\documentclass[aps,pra,twocolumn,amsmath,amssymb,showpacs,floatfix,superscriptaddress]{revtex4-2}

\usepackage{times}
\usepackage[utf8x]{inputenc}
\usepackage[english]{babel}
\usepackage[T1]{fontenc}
\usepackage{lmodern}
\usepackage{amssymb}
\usepackage{amsmath}
\usepackage{amsthm}
\usepackage[pdftex]{graphicx}
\usepackage{epstopdf}
\usepackage{hyperref}
\usepackage{braket}
\usepackage{comment}
\usepackage[bottom]{footmisc}
\usepackage{dcolumn}
\usepackage{bm}
\usepackage{color}
\usepackage{xcolor}
\usepackage{relsize,dsfont,mathrsfs,empheq,verbatim,upgreek}
\usepackage{etoolbox}
\usepackage[caption = false]{subfig}

\usepackage{enumerate}

\newcommand{\mathd}{\mathrm{d}}
\newcommand{\Tr}{\mathrm{Tr}}
\newcommand{\Lt}{\mathcal{L}}

\newcommand{\Dt}{\mathcal{D}}

\begin{document}


\title{Fluctuations of path-dependent thermodynamic quantities in open quantum systems via two-point system-only measurements
}


\author{Alessandra Colla}
\email{alessandra.colla@physik.uni-freiburg.de}
\affiliation{Institute of Physics, University of Freiburg, Hermann-Herder-Stra{\ss}e 3, D-79104 Freiburg, Germany}
\affiliation{Dipartimento di Fisica ``Aldo Pontremoli'', Universit\`a degli Studi di Milano, Via Celoria 16, I-20133 Milan, Italy}
\affiliation{INFN, Sezione di Milano, Via Celoria 16, I-20133 Milan, Italy}

\author{Andrea Smirne}
\affiliation{Dipartimento di Fisica ``Aldo Pontremoli'', Universit\`a degli Studi di Milano, Via Celoria 16, I-20133 Milan, Italy}
\affiliation{INFN, Sezione di Milano, Via Celoria 16, I-20133 Milan, Italy}

\author{Heinz-Peter Breuer}
\affiliation{Institute of Physics, University of Freiburg, Hermann-Herder-Stra{\ss}e 3, D-79104 Freiburg, Germany}
\affiliation{EUCOR Centre for Quantum Science and Quantum Computing, University of Freiburg, Hermann-Herder-Str. 3, D-79104 Freiburg, Germany}

\author{Bassano Vacchini}
\affiliation{Dipartimento di Fisica ``Aldo Pontremoli'', Universit\`a degli Studi di Milano, Via Celoria 16, I-20133 Milan, Italy}
\affiliation{INFN, Sezione di Milano, Via Celoria 16, I-20133 Milan, Italy}

\begin{abstract}
We propose a method to evaluate general thermodynamic fluctuations in open quantum systems, based on performing a two-point measurement scheme on the system using dynamics-dependent thermodynamic observables. Our approach allows one to obtain exact equalities for fluctuations of path-dependent thermodynamic quantities such as work and heat, and to isolate correction factors to Jarzynski's equality, requiring only access to the system degrees of freedom. This framework is flexible and can be applied to the limiting case of closed systems, recovering previous, yet seemingly contradictory, results from the literature. Moreover, the formalism admits a straightforward extension to strongly coupled open quantum systems. We investigate the effect of specific dynamical classes on the fluctuation relations, and show that the pure decoherence case is particularly special, as it deterministically does not contain any heat contribution and thus constitutes a class of open system dynamics for which the Jarzynski equality for work fluctuations is identically true at any coupling strength. Finally, we look explicitly at the shape and size of the correction factors to Jarzynski's equality for a qubit undergoing phase covariant dynamics, both in the weakly-coupled regime and in the deep non-Markovian regime.
\end{abstract}

\date{\today}

\maketitle

\section{Introduction}
Fluctuation theorems, such as the Jarzynski equality~\cite{Jarzynski1997} and the Crooks fluctuation theorem~\cite{Crooks1999}, are among the most celebrated results in classical nonequilibrium thermodynamics. They provide exact relations that constrain the statistics of nonequilibrium processes, establishing a connection between equilibrium thermodynamic properties and statistical fluctuations generated under external driving.

Several efforts have been made since to extend the power of these relations to the quantum domain \cite{Campisi2011,Aberg2018,Esposito2009a}. For closed quantum systems -- namely, systems evolving under the unitary evolution generated by a parameter-dependent Hamiltonian -- the \textit{two-point measurement scheme} (TPMS) \cite{Kurchan2001,Talkner2007a,Talkner2007b,Talkner2008} represents one of the most successful frameworks and yields a quantum version of the Jarzynski equality when applied to initial states that are diagonal in the energy eigenbasis of the initial Hamiltonian. 

The TPMS has been the subject of significant discussion. Most notably, the initial projective measurement destroys quantum coherences in the energy eigenbasis and is thus in general incompatible with mean energy variation estimates~\cite{Perarnau-Llobet2017,Beyer2020,Beyer2022}. Moreover, the protocol requires detailed prior knowledge of the system Hamiltonian, in contrast to the classical case where the Jarzynski equality also holds predictive power for equilibrium free energy differences \cite{Beyer2025}. Nevertheless, the TPMS has proven remarkably successful in formulating quantum fluctuation relations and provides a clear and operationally well-defined framework, within which the Jarzynski equality can be interpreted as a relevant constraint on thermodynamic quantities both at and out of equilibrium.

Many realistic thermodynamic scenarios, however, involve open quantum systems interacting with environments such as heat baths. Both in classical and quantum systems, work fluctuations for open systems that are very weakly coupled to memoryless environments can be evaluated by sampling/measuring over the global system, given that the ratio of global Gibbs partition functions reduces to that of the system alone \cite{Jarzynski1997,Campisi2011}. Similar reasoning can be applied for strongly coupled open systems if the Hamiltonian of mean force is used in the definition of thermodynamic quantities in this regime \cite{Talkner2020}.

While these considerations provide a formal framework, they involve procedures that are generally unfeasible in practice, such as measuring the global system, which includes potentially large, complex environments with many degrees of freedom. Furthermore, such schemes do not account for work contributions induced by the environment on the system during the process. Consequently, standard fluctuation theorem approaches fail to provide an operational description of work fluctuations for open systems.

So far, an accepted procedure for evaluating work fluctuations in open quantum systems accessing only the system degrees of freedom has been lacking \cite{Davoudi2025}. 
Some proposals have been put forth for weakly-coupled open quantum systems using dynamical maps and a TPMS \cite{Goold2021} or trajectories of the unravelings of time-local master equations \cite{Esposito2006,Leggio2013}, including a recent operational scheme providing bounds (rather than equalities) of predictive power \cite{Beyer2025}. Furthermore, alternative descriptions for open systems have been formulated using quasiprobability distributions \cite{Kwon2019,Gherardini2024,Pezzutto2025}.
These approaches are however only valid in specific Markovian regimes or bath modelings, and can rarely clearly distinguish between energy and work fluctuations when heat exchange with the environment is present.
Furthermore, even when accounting for strong coupling scenarios \cite{Sone2020}, they typically do not account for renormalization effects arising from strong coupling and non-Markovian environments.

In this work, we propose a flexible procedure, based on particular definitions of thermodynamic observables and on the TPMS, that allows one to obtain exact equalities for fluctuations of thermodynamic quantities in open quantum systems, and requires only \begin{enumerate}
    \item projective measurements on the reduced system;
    \item knowledge of the reduced system, and none of the environment.
\end{enumerate}
With the second point, we specifically assume that one knows the dynamical map of the system at all times of the protocol (and thus its Hamiltonian) as well as its initial state's eigenbasis. As such, this framework is not intended to provide predictive power over equilibrium quantities. It is however an operational framework, as the measurements and the knowledge required are achievable within the current capabilities of high-precision quantum platforms \cite{Lvovsky2009,Gebhart2023}. Its flexibility also allows us to address several critical points in the literature; in particular, the method proposed in this work is able to:
\begin{enumerate}
    \item distinguish between work, internal energy, and heat fluctuations;
    \item avoid the destruction of initial coherences, and thus reliably recover average quantities;
    \item provide exact equalities that recover known results in limiting cases, in particular Jarzynski's for closed quantum systems;
    \item explicitly isolate correction factors to Jarzynski's equality in open quantum systems;
    \item assess thermodynamic fluctuations for arbitrary open systems, including strong-coupling regimes and non-Markovian dynamics, taking into account emergent contributions due to the interaction with the environment;
    \item encompass different, and at times competing approaches in the literature, such as \cite{Allahverdyan2005,Kurchan2001,Talkner2007a,Talkner2007b,Talkner2008,Goold2021}.
\end{enumerate}

The framework we propose relies on the definition of open-system thermodynamic observables, i.e., Hermitian operators acting on the system Hilbert space. These observables will be used within a two-point measurement scheme (which can, for certain choices, reduce to a one-point measurement scheme) and are only constrained by average thermodynamic quantities. They are thus not unique, and the equalities found via this scheme do depend on the specific choice of observable. This is the reason why different approaches in the literature can coexist within our framework, which then ties the statistical fluctuation of thermodynamic quantities to the specific measurement protocol used. Moreover, our approach extends to generic regimes, and it avoids the use of sequential measurements in the course of the evolution, which would generally modify the dynamics -- including, in the limit of frequent monitoring, the freezing of evolution via the quantum Zeno effect -- and thus the thermodynamics of the reduced quantities of interest.

The remainder of the paper is organized as follows. In Sec.~\ref{sec:fluctuations-wc} we introduce our approach for evaluating thermodynamic fluctuations for open quantum systems weakly-coupled to heat baths, and find modifications to Jarzynski's equality for work fluctuations. There, we also emphasize the flexibility of the framework by showing how it includes different methods for work fluctuations in closed systems, and by treating systems with initial coherences. In Sec.~\ref{sec:non-Markov-extension} we show how the formalism can be extended to include renormalization effects in strongly-coupled and non-Markovian regimes, by employing the minimal dissipation definitions \cite{Colla2022} of thermodynamic quantities. In Sec.~\ref{sec:classes} we investigate the properties of the fluctuations relations we derived for specific classes of open system dynamics, such as pure decoherence and phase-covariant dynamics. Lastly, we show in Sec.~\ref{sec:example} quantitative results for a qubit undergoing phase-covariant dynamics, both in the weakly-coupled regime and in the deep non-Markovian regime.

\section{Energy, heat and work fluctuations for weakly coupled open quantum systems} \label{sec:fluctuations-wc}
In this section, we introduce our approach to evaluate thermodynamic fluctuations for the system of interest. Furthermore, we apply it to assess work fluctuations in closed systems with initial coherences, as well as internal energy, work and heat fluctuations in weakly-coupled open quantum systems. 

\subsection{General assumptions and the reduced TPMS} \label{sec:fluctuations-TPMS}

Let us first present the general assumptions behind the framework for evaluating thermodynamic fluctuations and recall the mechanism of a TPMS, in particular applied to arbitrary two-point observables related to an open-system quantity of interest.

We assume that the system is in contact with the environment and that some driving (work protocol) is performed on the system by an external agent, which has the effect of changing external parameters in the system Hamiltonian. This implies that we can write the bare system Hamiltonian with an explicit time-dependence; the total system thus evolves under the following Hamiltonian, comprised of system, environment, and interaction contributions:
\begin{eqnarray}
    H_{SE}(t) = H_S(t) + H_E + H_I \; .
\end{eqnarray}

Because of our focus on open system scenarios and techniques, let us first assume factorizing initial conditions $\rho_{SE}(0) = \rho_S(0)\otimes \rho_E(0)$, where the environment is assumed to start in a thermal state at inverse temperature $\beta$, $\rho_E(0)= e^{-\beta H_E}/Z_E$, with $Z_E= \Tr\{e^{-\beta H_E}\}$. For most of our discussion, we want to start in a situation as close as possible to an equilibrium scenario from the point of view of the open system. We assume that a local equilibrium point for the system is given by a Gibbs state with respect to the system Hamiltonian ${H}_S(0)$. We therefore assume, unless otherwise stated, that the initial state is given by
\begin{equation}
\rho_S(0) = \rho_S^G(0) = \frac{e^{-\beta {H}_S(0)}}{Z_S(0)}
\end{equation}
with $Z_S(0) = \Tr\{ e^{-\beta {H}_S(0)}\}$.
Note, however, that our scheme is also suitable for non-equilibrium initial states, see Sec.~\ref{sec:closed-coherences}.

The dynamics of the reduced system of interest is then obtained by tracing out the degrees of freedom of the environment after evolving the total system via $\mathcal{U}_t=\mathcal{T}e^{-i \int_0^t d\tau H_{SE}(\tau)}$, leading to the dynamical map
\begin{eqnarray}\label{eq:dynamical_map}
    \Phi_t[\rho_S(0)] = \Tr_E\{\mathcal{U}_t\rho_S(0)\otimes \rho_E(0)\mathcal{U}_t^\dag\} \;,
\end{eqnarray}
which is a linear completely positive and trace preserving \cite{Breuer2002,Rivas2011,Vacchini2024} superoperator propagating any initial system state $\rho_S(0)$.

Throughout the evolution, we can define a nonequilibrium free energy in analogy to macroscopic thermodynamics, using the von Neumann entropy $S(\rho_S(t))$ evaluated at any point in time of the open system evolution:
\begin{equation}\label{eq:neq-free-energy}
F_S(t) := U_S(t) - {1 \over \beta} S(\rho_S(t)) \; ,
\end{equation}
where $U_S(t)=\Tr\{H_S(t) \rho_S(t)\}$ is the system's nonequilibrium internal energy. If the system is initially in a Gibbs state, we find that its initial free energy is such that $F_S(0) =- {1 \over \beta} \ln (Z_S(0))$, in agreement with equilibrium scenarios.

If we now let the system evolve with the bath until an arbitrary time $t > 0$, the system will have in general evolved outside of an instantaneous equilibrium state $\rho_S(t) \neq \rho_S^G(t)$, and thus has a specific value of nonequilibrium free energy $F_S(t)$ that depends on the actual state of the system. However, in line with the original Jarzynski equality, it is useful to look at thermodynamic quantities of a system which is in equilibrium \emph{at the same conditions} as the actual system at time $t$. Namely, we define ``equilibrium counterparts'' of the usual quantities -- internal and free energy, entropy etc. -- by substituting the instantaneous Gibbs state $\rho_S^G(t)$ as their argument. We denote these equilibrium quantities with the same symbols as the nonequilibrium ones, but with an overhead bar (e.g. $\overline{U}_S(t)$, $\overline{S}_S(t)$, etc.). In particular, we find the equilibrium free energy as
\begin{align} \label{eq:free-energy-gibbs}
\overline{F}_S(t) = \Tr\{ {H}_S(t) \rho_S^G(t) \}  - {1\over \beta} S( \rho_S^G(t)) =  - {1 \over \beta} \ln (Z_S(t)) \; ,
\end{align}
with $Z_S(t) = \Tr\{ e^{-\beta {H}_S(t)} \}$ the partition function at time $t$. Then we directly find a relation between the change in equilibrium free energy $\overline{F}_S(t)$ of the open system and the ratio of partition functions:
\begin{equation}\label{eq:deltafree-part-open}
e^{-\beta \Delta \overline{F}_S(t)} = e^{-\beta \overline{F}_S(t)} e^{\beta \overline{F}_S(0)} = \frac{Z_S(t)}{Z_S(0)} \; ,
\end{equation}
which is formally equivalent to what happens in the classical case. In a similar way, this relation will enter our open system thermodynamic fluctuations, which we will formulate within a two-point measurement scheme. 

\subsubsection*{Fluctuations of open system quantities within a TPMS}

Consider a quantity $\Delta X(t)$ associated to a process and written as the variation in time of the expectation value of a system observable $O_x(t)$ (any time-dependent self-adjoint operator acting on the Hilbert space of the open system). Namely:
\begin{equation}\label{eq:mean-observable}
\Delta X(t) = \Tr\{O_x(t)\rho_S(t)\} - \Tr\{O_x(0) \rho_S(0)\}\; .
\end{equation}
For certain quantities, $O_x(t)$ can be chosen independently from the 
dynamics; internal energy difference (or equivalently work, in a closed system) is a direct example of this, with $O_x(t) = H_S(t)$. 
Most often, however, $O_x(t)$ will depend on the details of the system dynamics, so that its expectation-value variation is not
an exact differential (we will use the notation $\delta X(t)$ instead of $\Delta X(t)$ in these cases); relevant examples are indeed work
and heat in generic open system scenarios, see Sec.~\ref{sec:work-heat-wc}.

Assuming the final time to be fixed, we can then define a random variable associated to the quantity of interest, $x$, whose statistical fluctuations are determined by the fact that the observable $O_x(t)$ is measured projectively at the start and at the end of the protocol. Given the time-dependent spectral decomposition of the observable $O_x$:
\begin{equation}
    O_x(t) = \sum_{n} o^{(x)}_n(t) \ket{n(t)}\!\bra{n(t)} \; ,
\end{equation}
the probability distribution of $x$ is given by:
\begin{align} \label{eq:probability-distribution}
    p({x}) = \sum_{n,m} p_{m|n}(t) p_n(0) \delta({x}-o^{(x)}_m(t)+o^{(x)}_n(0)) \; ,
\end{align}
with probabilities
\begin{align} 
    p_n(0) =& \bra{n(0)}\rho_S(0)\ket{n(0)} \; , \\
    p_{m|n}(t)= & \bra{m(t)} \Phi_t [\ket{n(0)}\!\bra{n(0)}] \ket{m(t)} \; ,
\end{align}
where $\Phi_t$ is the open system's dynamical map, eq.~\eqref{eq:dynamical_map}. Notice that, since system and environment are assumed uncorrelated at time $t=0$, the dynamical map is not affected by the first measurement on the system. 

It is important to notice that to obtain a proper statistics of $x$ (i.e., one that at least reproduces $\Delta X(t)$ as the average), one must only consider initial system states that have no coherences in the eigenbasis of $O_x(0)$. If not, the destruction of coherences due to the initial measurement will skew the result of $\braket{x}$.
One can then analyze fluctuations of $x$ by computing its higher moments or generating functions. Let any of them be described by the smooth function $f(x)$, then we denote
\begin{align}\label{eq:average-f}
    \braket{f(x)}:=\int d {x} f({x}) p({x}) \; .
\end{align}
In the next sections, we will use this to recall the relation for the fluctuations of the open system internal energy, and study the ones of open system work and heat exchange. 

Since we allow the observable $O_x(t)$ to depend on the system dynamics (the ``path'' from $0$ to $t$), we are gifted with the fact that there is no unique time-dependent observable associated to the average value $\Delta X(t)$. Indeed, any other choice
\begin{equation}\label{eq:initial-condition}
    O'_x(t) = O_x(t) - (\Phi_t^{-1})^\dag[O_x(0) - O_x'(0)] \; ,
\end{equation}
with $O_x'(0)$ an arbitrary observable, gives the same mean value $\Delta X(t)$, see eq.~\eqref{eq:mean-observable}, for any initial state of the system. This means that the initial conditions on $O_x(t)$, and thus the first measurement, can in principle be chosen arbitrarily. This can be useful when trying to measure the statistics of a quantity for generic initial states of the system: by adjusting $O_x'(0)$ such that the initial measurement does not destroy any system coherences, we are able to faithfully reproduce expectation values. The tradeoff, however, is that different choices of observables will result in different measurement protocols, as well as different statistics for the observed quantity.

\subsection{Closed quantum system with initial coherences}\label{sec:closed-coherences}
Before we focus on thermodynamic fluctuations in open systems, let us show how to exploit the freedom in choice of observables to obtain a work fluctuation identity for any closed system, including when coherences with respect to the system Hamiltonian are present at the initial time. 

Indeed, the standard procedure of measuring the system Hamiltonian at initial and final times forces one to consider only initial system states that are diagonal in the eigenbasis of $H(0)$. 
Several efforts have been made in the literature to remedy this limitation \cite{Allahverdyan2014,Francica2020,Micadei2020,Levy2020,Diaz2020,Gherardini2021,Pei2023,Francica2024}, all of them departing from the standard TPMS.
Exploiting the freedom in the choice of the observable initial condition expressed by eq.~\eqref{eq:initial-condition}, we can instead still consider a TPMS in this scenario by choosing:
\begin{equation}
    O_w(0) = -\frac{1}{\beta}\ln[\rho(0)] \; ,
\end{equation}
where the $\ln$ function is intended to give zero for any zero eigenvalue of $\rho(0)$.

While this choice is perfectly valid for any value of $\beta$, it is useful to first settle on an appropriate value for it. Our goal in this section is to obtain a relation as close as possible to Jarzynski's equality, and such that it leads to it in the limit where the initial system state is a Gibbs state with respect to $H(0)$, for some temperature $\beta$. Therefore, we are looking for a value $\beta$ such that $\rho(0)$ is as close as possible to $e^{-\beta H(0)}/Z(0)$. Among the many distinguishability measures available (distances, divergences, etc...), a significant one to minimize here is the relative entropy between the two states. This is equivalent to asking that the system state has the same energy as its Gibbs counterpart \cite{Strasberg2021}, and is also such that the quantity
\begin{equation}
    \beta(F(0)-\overline{F}(0)) 
\end{equation}
is minimal. 

As we have fixed the temperature, we can also define a temperature-dependent operator $H^*_\beta$ such that 
\begin{equation}
    \rho(0)= \frac{e^{-\beta H^*_\beta}}{Z(0)} \; ,
\end{equation}
which is uniquely fixed by setting the partition function equal to the one given by $\Tr\{e^{-\beta H(0)}\}$. Then, the initial measurement is made with respect to the observable:
\begin{equation}
    O_w(0) = H^*_\beta +\frac{1}{\beta}\ln[Z(0)] \; ,
\end{equation}
while we define the operator 
\begin{equation}
    \xi_\beta = H^*_\beta - H(0) \; ,
\end{equation}
which encodes the deviation of the initial state from equilibrium. In general, it might contain both diagonal (departure from equilibrium distribution) and non-diagonal (coherences) contributions. 

Because of our choice of $O_w(0)$, to capture the statistics of work (or, identically, internal energy variation) we have to perform the second measurement at time $t$ of the observable
\begin{align}\nonumber
    O_w(t) =& H(t) - U_t[H(0) - O_w(0)]U_t^\dag \\ =& H(t) +  U_t \xi_\beta U_t^\dag  +\frac{1}{\beta}\ln[Z(0)]\; .
\end{align}
This expression ensures that the average work is correctly captured by the protocol, i.e. that eq.~\eqref{eq:mean-observable} is satisfied.

By now choosing $f(w)= e^{-\beta w}$ to study its fluctuations via the two-point measurement scheme, we obtain
\begin{equation}
    \braket{e^{-\beta w}} = \Tr\{ e^{-\beta O_w(t)} \} \; .
\end{equation}
This provides an exact equality for work fluctuations in the closed system at hand, for our chosen measurement protocol, which applies
even when there are initial coherences with respect to $H(0)$. While its interpretation is not immediately transparent, we can provide a more understandable upper bound for the fluctuations.
By employing the Golden-Thompson inequality \cite{Golden1965,Thompson1965}
\begin{equation}
    \Tr\{ e^{A+ B} \}\leqslant  \Tr\{ e^{A}e^{B} \}
\end{equation}
and the bound $\Tr\{A B\} \leqslant \lambda_{\mathrm{max}}\{A\}\Tr\{ B\} $ (the latter valid for positive semi-definite operators $A$ and $B$), where $\lambda_{\mathrm{min}}\{\cdot\}$ and $\lambda_{\mathrm{max}}\{\cdot\}$ denote the extreme eigenvalues of an operator, the work fluctuations can be bounded via:
\begin{align}\nonumber
    \braket{e^{-\beta w}} =&\frac{1}{Z(0)}\Tr\{ e^{-\beta (H(t) +  U_t \xi_\beta U_t^\dag)} \} \\  \leqslant &  \frac{1}{Z(0)}\Tr\{ e^{-\beta H(t)} e^{-\beta U_t \xi_\beta U_t^\dag} \}\\ \leqslant& \frac{Z(t)}{Z(0)} \lambda_{\mathrm{max}}\{e^{-\beta \xi_\beta } \}= e^{-\beta \Delta \overline{F}(t)} e^{-\beta \lambda_{\mathrm{min}}\{\xi_\beta\}} \; . \nonumber
\end{align} 

It shows that work fluctuations are bounded by a modification of the Jarzynski contribution, where the correction factor is explicitly related to the deviation of the initial system state from equilibrium via the minimum eigenvalue of $\xi_\beta$. Notice that this bound is saturated, and naturally gives the Jarzynski equality, whenever the initial system state starts in a Gibbs state.

Moreover, the above gives a clear bound to the average dissipated work via Jensen's inequality
\begin{align}
\label{eq:ribound}
    \braket{w} - \Delta \overline{F} \geqslant \lambda_{\mathrm{min}}\{\xi_\beta\} \; ,
\end{align}
which is valid for all protocol durations, only depends on the initial state of the system (or rather, its deviation from an equilibrium state), and leads to the standard second law for $\lambda_{\mathrm{min}}\{\xi_\beta\}=0$.

We remark that, on a first look, this approach might seem to defy the no-go theorem \cite{Perarnau-Llobet2017}, because it satisfies both the requests that the average work is found for all initial states and that correct Jarzynski is obtained for an initial Gibbs. However, the no-go theorem assumes the measurements to not depend explicitly on the initial state. Instead, this approach clearly does, with no contradiction with \cite{Perarnau-Llobet2017}.

\subsection{Internal energy fluctuations}
We can now proceed with evaluating thermodynamic fluctuations for open quantum systems. First, we recover the internal energy fluctuations of the open system already derived in Ref.~\cite{Goold2021}. This can be done by choosing $O_u(t)={H}_S(t)$. By denoting $u$ its associated variable and choosing, inspired by Jarzynski's equality, the function $f(u)=e^{-\beta u}$, we find via~\eqref{eq:probability-distribution}-\eqref{eq:average-f}
\begin{align}\nonumber
\langle e^{-\beta u} \rangle
=& \sum_{m,n} \frac{e^{-\beta e_m(t)}e^{+\beta e_n(0)}}{Z_S(0)} \times \\
& \bra{m(t)} \Phi_t[e^{-\beta e_n(0)}\ket{n(0)}\!\!\bra{n(0)}]\ket{m(t)} \nonumber \\ 
=&  \frac{1}{Z_S(0)} \Tr \{ e^{-\beta H_S(t)} \Phi_t [\mathbb{I}] \} \;,
\end{align}
where we denoted with $e_n(0)$, $e_m(t)$ the eigenvalues of $H_S(0)$ and $H_S(t)$, respectively, and $\ket{n(0)}$, $\ket{m(t)}$ their eigenstates. Introducing the partition function at time $t$ both at the numerator and denominator, we obtain
\begin{align}\nonumber
\langle e^{-\beta u} \rangle
=&  \frac{Z_S(t)}{Z_S(0)} \Tr \{ \rho_S^G(t) \Phi_t [\mathbb{I}] \}  \\ \label{eq:J-equality-U}
=&  \Lambda^u_S(t) e^{-\beta \Delta \bar{F}_S(t)}  \;,
\end{align}
where we have introduced the factor \begin{equation}
    \Lambda^u_S(t) :=  \Tr \{ \rho_S^G(t) \Phi_t [\mathbb{I}] \} \;,
\end{equation} which appears as an open system contribution to a modified Jarzynski equality. Eq.~\eqref{eq:J-equality-U} is a general relation for the fluctuations of the internal energy of the open system; one can directly recover Jarzynski's equality for closed systems from it by substituting ${H}_S(t)$ with ${H}(t)$ -- the closed system Hamiltonian -- and the dynamical map $\Phi_t$ with a unitary evolution $\mathcal{U}_t$, for which $\Lambda^u_S(t) = 1$. 

As noticed in \cite{Goold2021}, for open systems with unital dynamical map ($\Phi_t[\mathbb{I}] = \mathbb{I}$) the factor $\Lambda^u_S(t)$ gives the identity, and we obtain exactly Jarzynski's result on the right-hand-side of eq. \eqref{eq:J-equality-U}.

While the result in the general case is similar to a Jarzynski equality, the term $\Lambda^u_S(t)$ provides a correction due to open system dynamics.
We can interpret the term in different ways by massaging its expression. In particular, we can write
\begin{equation}\label{eq:lambda_u}
 \Lambda^u_S(t) = \Tr\{ \Phi^{\dagger}_t [\rho_S^G(t)]\} \; ,
\end{equation}
where $\Phi^{\dagger}_t$ denotes the adjoint of the dynamical map with respect to the Hilbert-Schmidt scalar product. Viewing this as a Heisenberg picture, this shows how the sum of probabilities associated to the instantaneous Gibbs state is changed under the action of the adjoint dynamical map.  If the total probability of $\rho_S^{G}(t)$ is conserved in the Heisenberg picture, then we obtain an equality for the fluctuations of internal energy as for a closed system.

Alternatively, by dividing and multiplying the expression by the open system's Hilbert space dimension $d$, we can write
\begin{equation}
 \Lambda^u_S(t) = d\, \Tr\{ \rho_S^G(t)  \Phi_t [ \mathbb{I}/d ]\}  ,
\end{equation}
meaning that the term $\Lambda^u_S(t)$ is proportional (with factor of Hilbert space dimension) to the Hilbert-Schmidt scalar product between the instantaneous Gibbs state $\rho_S^G(t)$ and the evolution, up to time $t$ and under the dynamical map $\Phi_t$, of the maximally mixed state of the system. The larger the overlap, i.e. the closer the instantaneous Gibbs state looks to an evolved infinite temperature state, the higher are the fluctuations of internal energy. 

Lastly, we can also bound the coefficient by
\begin{equation}
 \Lambda^u_S(t) \leqslant \lambda_{\mathrm{max}}\{\Phi_t[\mathbb{I}]\} \; ,
\end{equation}
which takes the meaning of a correction factor that is due to the non-unitality of the dynamical map. The bound is indeed saturated for unital maps.

\subsection{Work and heat fluctuations}\label{sec:work-heat-wc}
In this section, we define the observables related to the work and heat exchanged during a driving process in an open quantum system. Unless otherwise specified, the denominations ``heat'' and ``work'' are intended to mean their average value, as often done in the literature.
We recall the definition of their expected average value according to weak-coupling quantum thermodynamics \cite{Alicki1979,Kosloff2013}:
\begin{align}\label{eq:work-wc}
    \delta W_S(t) =& \int_0^t d\tau \Tr\{\dot{{H}}_S(\tau) \rho_S(\tau)\} \; , \\ \label{eq:heat-wc}
    \delta Q_S(t) =& \int_0^t d\tau \Tr\{{H}_S(\tau) \dot{\rho}_S(\tau)\} \; . 
\end{align}
We then use them within the TPMS to obtain modified equalities for their fluctuations. 

\subsubsection{Work and heat observables}\label{sec:wh-observables}
An appropriate observable for work exchange should be a self-adjoint operator such that the difference of its expectation value at time $t$ and time $0$ gives exactly the work done on (or by) the system between these two times. Namely, we search for an operator $O_w(t)$ such that
\begin{equation}\label{eq:FT-work-exchange-expectation}
\langle O_w(t) \rangle_t - \langle O_w(0)\rangle_0 = \delta W_S(t) \; ,
\end{equation}
where, compared with Eq.(\ref{eq:mean-observable}), we introduced the notation $\langle O \rangle_t = \Tr\left\{O \rho_S(t)\right\}$.
Since work is a path-dependent quantity whenever there is heat exchanged with the system (i.e., it depends on the dynamics of the system through the whole process), its observable naturally depends on the system dynamics.

We can construct this operator by defining the inverse propagator $\Phi_{\tau,t}$, with $\tau\leqslant t$, which relates the state of the system at a past time $\tau$ to the state at time $t$, i.e. the superoperator for which
\begin{equation}
{\rho}_S(\tau) = \Phi_{\tau,t} [\rho_S(t)] \; .
\end{equation}
In terms of the dynamical map for the system, the inverse propagator is given by $\Phi_{\tau,t} = {\Phi}_{\tau} \circ \Phi_t^{-1}$.  
Then, using its dual and recalling the freedom in Eq.(\ref{eq:initial-condition}), a proper work observable can be written as
\begin{equation}
O_w(t) := \int_0^t \mathd \tau \Phi_{\tau,t}^{\dagger}[\dot{{H}}_S(\tau)] + (\Phi_t^{-1})^\dag[O_w(0)]\;, 
\end{equation}
for an arbitrary operator $O_w(0)$.

By defining the time-local generator of the dynamics $\Lt_t=\dot{\Phi}_{t} \circ \Phi_t^{-1} $, and noting that $\Lt_{\tau}\Phi_{\tau,t} = \dot{\Phi}_{\tau} \circ \Phi_t^{-1}$, the operator above can be written equivalently as
\begin{equation}
O_w(t) := {H}_S(t) - \mathfrak{P}(t)+ (\Phi_t^{-1})^\dag[O_w(0)-H_S(0)] \;,
\end{equation}
where $\mathfrak{P}(t)$ singles out the contribution due to the fact that work in the open system is path-dependent, and it is given by
\begin{equation}\label{eq:P_wc}
\mathfrak{P}(t)=\int_0^t \mathd \tau 
\Phi_{\tau,t}^{\dagger}\circ\Dt_\tau^{\dagger}[{H}_S(\tau)] \; .
\end{equation}
Here, $\Dt_t[\cdot] = \Lt_t[\cdot] +i[H_S(t),\cdot]$ is the dissipator part of the time-local generator of the dynamics.
Indeed, $\mathfrak{P}(t)$ is identically zero for unitary evolutions (closed systems) and, more generally, for any dynamics such that
$\Dt_t^{\dagger}[{H}_S(t)]=0$, see, e.g., pure decoherence (sec.~\ref{sec:dephasing}).

A similar procedure can be followed to define a heat observable so that the difference of expectation values before and after the protocol gives the heat exchanged by the open system, eq.~\eqref{eq:heat-wc}, 
finding
\begin{equation}
O_q(t) := \mathfrak{P}(t)+ (\Phi_t^{-1})^\dag[O_q(0)]\; .
\end{equation}

\subsubsection{Single-time vs. two-time measurement protocols}
The freedom in selecting the operators \( O_w(0) \) and \( O_q(0) \) arises from the fact that our definitions of the heat and work operators are constrained solely by the variations of their expectation values.
In fact, it allows us to devise qualitatively different schemes, characterized by, respectively, single-time and two-time measurement protocols, in this way providing a unified framework for different approaches in the literature.

First, we can indeed make the choice $O_w(0)=O_q(0)=0$, meaning that both work and heat can be evaluated via a single measurement at time $t$. A similar choice can be made by selecting instead the initial operators such that $O_w(t)=O_q(t)=0$, where the thermodynamic quantities are then evaluated via a single measurement at time $t=0$. In this second case, the request that the initial state of the systems be diagonal in the eigenbasis of the operators $O_w(0)$ and $O_q(0)$ is no longer necessary, since the absence of the second measurement is enough to ensure that the expectation value is correctly recovered. 

While we will mostly focus on two-point measurement schemes for the evaluation of work and its fluctuations, it is worth it to briefly consider the case of a single measurement at time $t=0$. With that choice, one has $\delta W_S(t)= -\Tr\{O_w(0) \rho_S(0)\}$. This scenario reduces to the approach proposed in \cite{Allahverdyan2005}, where the work operator $\Omega = U_t^\dag H(t) U_t - H(0)$ is defined for a closed system such that $\delta W(t)= \Tr\{\Omega \rho(0)\}$. Indeed, imposing $O_w(t)=0$ for a closed system we obtain $O_w(0)=-\Omega$. This shows how our scheme, which includes the freedom of choice of initial measurements, can bridge between the approach of a single work operator and that of an energetic two-point measurement scheme.

Indeed, a different choice can be made by asking that the variation of the work, heat and energy operators satisfy
\begin{equation}\label{eq:bal}
    O_w(t) - O_w(0) + O_q(t) - O_q(0) = H_{S}(t)-H_S(0).
\end{equation}
Such a balance is not strictly needed for the validity of the first law of thermodynamics at the level of the expectation values -- 
for which Eqs.(\ref{eq:work-wc}) and (\ref{eq:heat-wc}) are enough. 
However, Eq.(\ref{eq:bal}) can be motivated by thinking of the classical limit obtained by imposing that all operators are diagonal in the same basis,
and associating them with functions on the phase space, which should satisfy the first law of thermodynamics for every trajectory on it.
From here on, we will enforce Eq.(\ref{eq:bal}), by setting, in particular, $O_w(0)=H_S(0)$, that is, work is measured via a two-time protocol where the first measurement corresponds to the measurement of the system Hamiltonian,
while at time $t$ one measures
\begin{equation}
O_w(t) = {H}_S(t) - \mathfrak{P}(t)\;.
\end{equation}
Consequently, $O_q(0)=0$, i.e., heat is obtained via the one-time measurement of 
\begin{equation}\label{eq:Oq_P}
O_q(t) = \mathfrak{P}(t)\; .
\end{equation}

With the above we have constructed observables that give the right physical quantity when measured at time $t$ or at times $0$ and $t$, even if the quantity depends on what happens \emph{during} the time interval $[0,t]$. Of course, to be able to do that one needs to inject the theoretical knowledge of the ideal evolution of the system in that interval; nonetheless, statistical uncertainty and fluctuations can still arise from this ideal knowledge when the observable is actually measured.

\subsubsection{Open system work and heat equalities}
For any two-point observable such that $O_x(0)={H}_S(0)$, and with the open system initially in a Gibbs state $\rho_S(0)=\rho_S^G(0)$ as we assume, one has that 
\begin{align}
    p_n(0) = e^{-\beta x_n(0)} \; .
\end{align}
Then, performing the two-point measurement scheme for this observable and looking at its fluctuations using $f(x)= e^{-\beta x}$ gives
\begin{align} \nonumber 
\langle e^{-\beta x}\rangle = \frac{Z_S(t)}{Z_S(0)} \Tr \left\{  \frac{e^{-\beta O_x(t)}}{Z_S(t)}\Phi_t[\mathbb{I}] \right\}\; .
\end{align}
With the notation $\Lambda_S^x(t):= \Tr \left\{  \frac{e^{-\beta O_x(t)}}{Z_S(t)}\Phi_t[\mathbb{I}] \right\}$ for the correction factor, we find a modified Jarzynski equality for open quantum systems at arbitrary couplings, by using $x=w$:
\begin{equation}\label{eq:J-arbitrary}
\langle e^{-\beta w}\rangle = \Lambda_S^w(t) e^{-\beta \Delta \bar{F}_S(t)} ,
\end{equation}
where the correction factor reads
\begin{equation}\label{eq:lambda_w}
    \Lambda_S^w(t)= \Tr \left\{  \frac{e^{-\beta O_w(t)}}{Z_S(t)}\Phi_t[\mathbb{I}] \right\}\; .
\end{equation}
The above coefficient is now not only a correction factor due to open system dynamics, but also due to the fact that the quantity is path-dependent. Even if the dynamical map is unital, there is no guarantee that the exponential of the work observable will have trace equal to the partition function at time $t$, and thus that the correction factor will be equal to one. For the limit in which there is no heat exchange during the whole process, such that $O_w(t)={H}_S(t)$, the work fluctuations are equal to those for internal energy \eqref{eq:J-equality-U}, and the conditions for the correction factor to be equal to one are the same as explained in the previous section. An example of this is given by pure decoherence dynamics, see Sec.~\ref{sec:dephasing}.

Similarly to what we did in Sec.~\ref{sec:closed-coherences}, we can use basic inequalities for the trace of a product of matrices along with the Golden-Thompson inequality:
\begin{align} \nonumber
    &\frac{1}{Z_S(t)} \Tr\{ e^{-\beta(H_S(t)-\mathfrak{P}(t))}\}\lambda_{\mathrm{max}}\{\Phi_t[\mathbb{I}]\} \\ \nonumber
    &\leqslant\frac{1}{Z_S(t)}   \Tr\{ e^{-\beta H_S(t)} e^{\beta \mathfrak{P}(t)}\}\lambda_{\mathrm{max}}\{\Phi_t[\mathbb{I}]\} \\
    &\leqslant e^{\beta \lambda_{\mathrm{max}}\{\mathfrak{P}(t)\}}  \lambda_{\mathrm{max}}\{\Phi_t[\mathbb{I}]\} \; ,
\end{align}
where the first inequality is saturated for $[H_S(t),\mathfrak{P}(t)]=0$ (as is the case for phase-covariant dynamics, see Sec.~\ref{sec:phase-covariant}). It leads to an interpretable upper bound ($\mathcal{UB}$) for the correction factor to work fluctuations, $\Lambda_S^w(t)\leqslant \mathcal{UB}[\Lambda_S^w(t)]$, with
\begin{equation}\label{eq:lambdaw_wc_bound}
    \mathcal{UB}[\Lambda_S^w(t)] = e^{\beta \lambda_{\mathrm{max}}\{\mathfrak{P}(t)\}}  \lambda_{\mathrm{max}}\{\Phi_t[\mathbb{I}]\} \; .
\end{equation}
This bound clearly distinguishes between a contribution due to non-unitality of the dynamical map (the same found for internal energy fluctuations, saturated for unital maps) and one due to the fact that the system exchanges heat with the environment (saturated when the dynamics is such that there is no heat exchange for any initial state). 

By applying Jensen's inequality, we can therefore also obtain a bound for the average dissipated work in an open quantum system:
\begin{equation}
    \braket{w}-\Delta \overline{F} \geqslant - \lambda_{\mathrm{max}}\{\mathfrak{P}(t)\} -\frac{1}{\beta}\ln(\lambda_{\mathrm{max}}\{\Phi_t[\mathbb{I}]\} ) \; .
\end{equation}
This bound now depends on the duration of the work protocol and extends the standard constraint $\braket{w}-\Delta \overline{F} \geqslant 0$ valid for closed systems.

Statistical fluctuations of heat can also be obtained via an analogous procedure, by choosing $x=q$. Because of our chosen initial conditions for the heat operator, eq.~\eqref{eq:Oq_P}, the TPMS in this case reduces to a one-point measurement performed at the end of the protocol. This gives
\begin{equation}
    \braket{e^{-\beta q}}= \Tr\{ e^{-\beta \mathfrak{P}(t)} \Phi_t[\rho_S(0)] \} \leqslant e^{-\beta \lambda_{\mathrm{min}}\{\mathfrak{P}(t)\}} \; . \label{eq:heat-fluctuations}
\end{equation}

It is crucial to remark that the construction of these procedures depends explicitly on the inverse dynamical map $\Phi_t^{-1}$, as the work and heat exchanged with the system require knowledge of the state at all intermediate times of the protocol, while the second measurement is performed only at the final time.
While this is formally well-defined as long as the map remains invertible, the magnitude of the inverse map may diverge as the system evolves. This typically occurs as the system approaches a fixed point, such as a thermal state, where distinct initial states are mapped to states with vanishingly small distinguishability, or more in general when at least two distant initial states are mapped into states with very small distinguishability. Notably, this divergence may not be apparent in the instantaneous generator $\Lt_t$, because the large component of the inverse is often compensated by the vanishing time-derivative of the map $\dot\Phi_t$. However, this mitigation mechanism might fail in our expression~\eqref{eq:P_wc}, as the inverse map and the time-derivative are considered at different times; for an explicit example, see eq.~\eqref{eq:q-pc}.

\section{Extension to arbitrary coupling and non-Markovian systems}\label{sec:non-Markov-extension}

In this section, we expand the framework illustrated above in order to account for arbitrary coupling strengths between the system and the bath, and to include the presence of memory effects. The main difficulty with understanding fluctuations of thermodynamic quantities in this regime is that the quantities themselves do not have a widely accepted definition. Indeed, many approaches have so far been proposed in the literature --  see, e.g., \cite{Weimer2008,Esposito2010,Alipour2016,Strasberg2017,Rivas2020,Alipour2021,Landi2021,Elouard2023} -- and may give non-compatible definitions for quantities like work and internal energy, for example \cite{Colla2025local,Picatoste2025}.
Moreover, some of these proposals are based on measuring the expectation value of the environmental Hamiltonian $H_E$, a task which is unfeasible in realistic scenarios.

In order to maintain the main goal set in this work to make use of information available through the open system dynamics only, a suitable candidate framework for the definition of thermodynamic quantities in arbitrary regimes is the one of minimal dissipation \cite{Colla2022}, as it shares the same goal. In Sec.~\ref{sec:minimal-dissipation} we recall the main definitions according to the minimal dissipation open system quantum thermodynamics, while in Sec.~\ref{sec:extension} we show how to use it to straightforwardly extend the methods of Sec.~\ref{sec:fluctuations-wc} to arbitrary coupling regimes.

\subsection{Minimal dissipation quantum thermodynamics}\label{sec:minimal-dissipation}

The framework proposed in \cite{Colla2022} allows one to define the principal thermodynamic quantities for an arbitrary open quantum system (in particular, possibly strongly coupled to a non-Markovian environment) solely by knowing the dynamical map $\Phi_t$ governing the open system dynamics. Whenever such a map is invertible \cite{Stelmachovic2001,Breuer2012,Hall2014}, it induces the existence of an exact time-local, or time-convolutionless (TCL), master equation for the reduced system
\begin{equation} \label{tcl-master}
 \frac{d}{dt}\rho_S(t) = \Lt_t[\rho_S(t)],
\end{equation}
through the construction of the time-local generator $\Lt_t=\dot{\Phi}_t \Phi_t^{-1}$. Such master equation, as we have also used and mentioned in Sec.~\ref{sec:work-heat-wc}, can always be written in the following general form \cite{Breuer2012,Hall2014}:
\begin{equation} \label{eq:generator-TCL}
\Lt_t[\rho_S(t)]= -\mathrm{i} \left[K_S(t),\rho_S\right] + {\mathcal{D}}_t[\rho_S(t)],
\end{equation}
with the dissipative part given by
\begin{equation} \label{Diss-part}
 {\mathcal{D}}_t [\rho_S] = \sum_{k}\gamma_{k}(t)\Big[L_{k}(t)
  \rho_S L_{k}^{\dag}(t) - \frac{1}{2}\big\{L_{k}^{\dag}(t)L_{k}(t),\rho_S\big\}\Big] \; .
\end{equation}
We emphasize that the above represents an exact treatment whenever the dynamical map is exact. In particular, it is valid in situations outside of the weak-coupling, Markovian regimes. In this case, the Hamiltonian $K_S(t)$ appearing in the coherent part of the generator is an effective Hamiltonian, possibly distinct from the bare microscopic system Hamiltonian $H_S(t)$. Furthermore, the dissipator rates $\gamma_k$, in contrast to the weak-coupling limit, may temporarily take on negative values. This means that the dynamical propagator at intermediate times may not be a completely positive (or even positive) map, which may signal the presence of non-Markovian behavior \cite{Breuer2009,Rivas2010,Hall2014,Wissmann2015,Breuer2016C,Chruscinski2022}.

The effective Hamiltonian $K_S(t)$ is in general affected by all kinds of non-Markovian and strong coupling effects due to the interaction with the environment, and is thus a prime candidate for the observable associated to the internal energy of the reduced system alone. However, the splitting of the master equation \eqref{eq:generator-TCL} into a coherent part and a dissipator is not unique, thus there is in principle no unique effective Hamiltonian \cite{Colla2022}. A preferential splitting proposed in \cite{Sorce2022} is based on a minimization procedure for the size of the dissipator (thus the term \textit{minimal dissipation}), and leads to the choice of the (unique) dissipator that is written in terms of traceless Lindblad operators. In turn, this specifies the unique Hamiltonian $K_S(t)$. This particular choice is convincing also because it corresponds to the natural splitting one obtains in the weak coupling regime from the microscopic derivation \cite{Breuer2002}; moreover, the renormalized transition frequencies of the open system predicted by this unique effective Hamiltonian have been recently observed in trapped ions \cite{Colla2024exp}.

Once the effective Hamiltonian is determined, the internal energy of the reduced system $S$ is defined as $U_S (t) = \Tr \{K_S(t)\rho_S(t)\}$, while heat and work are defined as
\begin{eqnarray}
 \delta W_S(t) &=& \int_0^t \mathd \tau \, \Tr \big\{ \dot{K}_S(\tau) \rho_S(\tau) \big\},  \label{work} \\
 \delta Q_S(t) &=& \int_0^t \mathd \tau \, \Tr \big\{ K_S(\tau) \dot{\rho}_S(\tau)  \big\}  \label{heat}
\end{eqnarray}
in agreement with a reduced-system first law
\begin{eqnarray}
\Delta U_S (t) = \delta W_S(t) + \delta Q_S(t) \; .
\end{eqnarray}

Notice that the only part of the evolution of the density matrix which affects heat in its definition is the dissipative part, $\delta Q_S(t) = \int_0^t \mathd \tau \, \Tr \big\{ K_S(\tau) \mathcal{D}_{\tau}[\rho_S(\tau)] \big\}$, similarly to the weak-coupling case. Moreover, the definition of work allows the environment to perform work on the reduced system, since $K_S$ can be time-dependent even if no explicit work protocol was imposed on the system, i.e. even if $H_S$ was taken to be time independent in the total Hamiltonian. This makes particular sense when thinking of environments starting in a coherent state, where emergent driving is seen to appear already at a first-order approximation in the coupling strength \cite{Colla2024roles} and has been associated with work also in other approaches \cite{Cavina2024}.

The study of thermodynamic properties according to the minimal dissipation approach has been conducted in many integrable models where one has access to the exact dynamical map. This includes the case of a qubit interacting with a single bosonic environmental mode -- such as Jaynes-Cummings and variations \cite{Colla2022,Seegebrecht2024} -- or a single environmental qubit \cite{Neves2025}, the Fano-Anderson model \cite{Picatoste2024,Colla2024roles,Colla2025local} and pure decoherence models \cite{Picatoste2025}. Furthermore, numerical methods for the simulation of open quantum system dynamics are also particularly suitable for the calculation of these quantities; this has been done, for example, for the finite-temperature spin-boson model \cite{Gatto2024} and the single impurity Anderson model \cite{Gatto2026}. Lastly, perturbative and recursive approaches are also available to find the dynamics as well as the effective Hamiltonian when an exact treatment is not possible \cite{Gasbarri2018,Colla2025k,Colla2025recursive}.

\subsection{Two-point thermodynamic fluctuations for arbitrary open systems} \label{sec:extension}

Let us now make use of the definitions for the thermodynamic quantities proposed in the minimal dissipation approach to extend the two-point measurement framework described in Sec.~\ref{sec:fluctuations-wc} and study their fluctuations outside of the weak-coupling regime. 

Because the minimal dissipation framework is formally very similar to the open-system, weak-coupling approach to quantum thermodynamics, the extension to non-Markovian open quantum system is straightforward, and amounts to simply considering the renormalized Hamiltonian $K_S(t)$ as the internal energy operator for the system instead of the bare Hamiltonian $H_S(t)$.

As before, we assume factorizing initial conditions $\rho_{SE}(0) = \rho_S(0)\otimes \rho_E(0)$. Moreover, we assume that a local equilibrium point for the system at initial times is given by a Gibbs state with respect to the renormalized system Hamiltonian ${K}_S(0)$. Notice, however, that in most standard cases the effective Hamiltonian starts unrenormalized, i.e. $K_S(0)=H_S(0)$. This is true whenever
\begin{equation}
    \Tr_E\{\tilde{H}(t)\rho_E(0)\} = 0 \; ,
\end{equation}
where $\tilde{H}(t)$ is the total Hamiltonian in the interaction picture, and which leads to a vanishing first order contribution to the generator expansion in orders of the coupling strength \cite{Rivas2011,Colla2025k,Colla2025recursive}. To account for the most general case, however, we assume that the initial state is given by
\begin{equation}
\rho_S(0) = \rho_S^G(0) = \frac{e^{-\beta {K}_S(0)}}{\Tr\{e^{-\beta {K}_S(0)}\}}\; .
\end{equation}

Now, the driving (work protocol) that is performed on the system can be done either by an external agent, via an explicit time-dependence of the bare Hamiltonian, or by the interaction with the environment, through an emergent time-dependence of the effective Hamiltonian. These two effects are not mutually exclusive, and are not additive; namely, the interaction with the environment can also be responsible for the renormalization of the external driving protocol \cite{Breuer1997,Colla2024roles}.

We again define the nonequilibrium free energy as in eq.~\eqref{eq:neq-free-energy}, but this time using the renormalized internal energy $U_S(t)=\Tr\{K_S(t) \rho_S(t)\}$. We also define its ``equilibrium counterpart'' by substituting the instantaneous renormalized Gibbs state $\rho_S^G(t)$ as its argument to find
\begin{align} \label{eq:free-energy-gibbs}
\overline{F}_S(t) = \Tr\{ {K}_S(t) \rho_S^G(t) \}  - {1\over \beta} S( \rho_S^G(t)) =  - {1 \over \beta} \ln (Z_S(t)) \; ,
\end{align}
with $Z_S(t) = \Tr\{ e^{-\beta {K}_S(t)} \}$ the partition function at time $t$. Because the structure is the same as before but with the effective Hamiltonian, we recover formally eq.~\eqref{eq:deltafree-part-open}.

The strategy to find the fluctuations of open system quantities through a TPMS remains identical as established in Sec.~\ref{sec:fluctuations-TPMS}. The conceptual difference is that the dynamical map describing the evolution of the system can now be non-Markovian and that the thermodynamic quantities are taking the renormalization of the system Hamiltonian into account.

We thus investigate internal energy fluctuations by choosing $O_u(t)={K}_S(t)$. Because everything is formally equivalent as before, we find
$\langle e^{-\beta u} \rangle = \Lambda^u_S(t) e^{-\beta \Delta \bar{F}_S(t)}$, with 
$\Lambda^u_S(t) :=  \Tr \{ \rho_S^G(t) \Phi_t [\mathbb{I}] \} $. This corresponds again to a modified Jarzynski equality, where open systems with unital dynamical map ($\Phi_t[\mathbb{I}] = \mathbb{I}$) have once more $\Lambda^u_S(t)=1$, and recover Jarzynski's result as long as one takes renormalization into account in the partition functions and in the definition of equilibrium free energy.

We can then construct the work and heat operators just as in Sec.~\ref{sec:work-heat-wc}, but substituting the renormalized Hamiltonian in each definition, to obtain
\begin{equation}
O_w(t) := {K}_S(t) - \mathfrak{P}(t) \;,
\end{equation}
where $\mathfrak{P}(t)$, and therefore the heat operator, is given by
\begin{equation}\label{eq:P_md}
\mathfrak{P}(t)= O_q(t) = \int_0^t \mathd \tau 
\Phi_{\tau,t}^{\dagger}\circ\Dt_\tau^{\dagger}[{K}_S(\tau)] \; .
\end{equation}

Following the TPMS for the work operator, we formally obtain again eq.~\eqref{eq:J-arbitrary}, with coefficient
\begin{equation}\label{eq:lambda_w_md}
\Lambda_S^w(t):= \Tr \left\{  \frac{e^{-\beta [K_S(t) - \mathfrak{P}(t)]}}{Z_S(t)}\Phi_t[\mathbb{I}] \right\} \; .
\end{equation}
Notice that an identical bound as eq.~\eqref{eq:lambdaw_wc_bound} also holds for this coefficient. Similarly for heat, we obtain again eq.~\eqref{eq:heat-fluctuations}.

\section{Specific dynamical classes}\label{sec:classes}
Before we look at some specific examples, it is useful to make a few general remarks based on whether the open system dynamics has some specific properties. We discuss the case of phase covariant dynamics and pure decoherence. In what follows, we assume driving on the system degrees of freedom where the Hamiltonian commutes with itself at all times, namely $[H_S(t),H_S(t')]=0$, $\forall t,\; t'$.

\subsection{Phase covariant dynamics}\label{sec:phase-covariant}

By phase covariant dynamics, we mean that the dynamical map $\Phi_t$ commutes with a one parameter unitary group $V_\lambda[\cdot]= e^{-i\lambda S}\cdot e^{i\lambda S}$ for some Hermitian operator $S$, which we assume commutes with ${H}_S(t)$. Then, due to this symmetry, all the superoperators involved in the dynamics have the following form \cite{Chruscinski2022}:
\begin{equation}\label{eq:pc-structure}
    \Psi[\cdot] = \sum_{jk}a_{jk}[\Psi] \ket{j}\!\bra{k}\cdot \ket{k}\!\bra{j}+ \sum_{j\neq k}b_{jk}[\Psi] \ket{j}\!\bra{j}\cdot \ket{k}\!\bra{k} \; ,
\end{equation}
where $\ket{j}$ are the eigenvectors of ${H}_S(t)$ \cite{Vacchini2010a}.
Therefore, any phase covariant map $\Psi$ of this kind is described by the 
set of coefficients $\{a_{jk}[\Psi]\}$ that can be conveniently identified with a matrix, and the set $\{b_{jk}[\Psi]\}$ defined only for $i\neq j$.
Considering the dynamical map $\Phi_t$ we can thus write
\begin{equation}\label{eq:dynamical_map_pc}
    \Phi_t [\cdot] = \sum_{jk}F(t)_{jk} \ket{j}\!\bra{k}\cdot \ket{k}\!\bra{j}+ \sum_{j\neq k}f_{jk}(t) \ket{j}\!\bra{j}\cdot \ket{k}\!\bra{k} \; ,
\end{equation}
where due to trace and Hermiticity preservation $F(t)$ is a real time-dependent matrix defined as
\begin{equation}
    F(t)_{jk}=a_{jk}[\Phi_t]
\end{equation}
and obeying $\sum_j F(t)_{jk}=1 \ \forall k$, while for $i\neq j$ we have introduced the time-dependent coefficients
\begin{equation}
    f_{jk}(t)=b_{jk}[\Phi_t]
\end{equation}
obeying $f_{jk}(t)=f^*_{kj}(t)$. 
In this notation the time derivative and inverse of the dynamical map are  described by
\begin{align}
    &a_{jk}[\dot\Phi_t] = \left(\dot{F}(t)\right)_{jk}\;, \quad  \quad \quad b_{jk}[\dot\Phi_t] = \dot{f}_{jk}(t) \\
    &a_{jk}[\Phi^{-1}_t] = \left({F^{-1}}(t)\right)_{jk} \; ,\quad b_{jk}[\Phi^{-1}_t] = \frac{1}{f_{jk}(t)}
\end{align}
which gives for the set of coefficients describing the time-local generator
\begin{align}
    &a_{jk}[\Lt_t] = \left(\dot{F}(t) F^{-1}(t)\right)_{jk}\;,  \quad \quad b_{jk}[\Lt_t] =  \frac{\dot{f}_{jk}(t)}{f_{jk}(t)} \; .
\end{align}

The effective Hamiltonian can be then computed via the following equation \cite{Colla2025k}:
\begin{align}\label{eq:K-jk}
K_S(t) = \frac{1}{2i d} \sum_{j,k=1}^d [|j\rangle\langle k|,\,\mathcal{L}_t[|k\rangle\langle j|]],
\end{align}
which is valid using any orthonormal basis ${\ket{j}}$. This leads to a Hamiltonian that remains diagonal in the original basis, namely
\begin{equation}
    K_S(t) = \sum_j k_j(t) \ket{j}\!\!\bra{j} \\  \label{eq:k-pc}
\end{equation}
with
\begin{equation}
    k_j(t)= -\frac{1}{d}\sum_k \text{Im}\left\{ \frac{\dot{f}_{jk}(t)}{f_{jk}(t)} \right\} \; ,
\end{equation}
where there is no contribution coming from the part of the generator described by the matrix associated to the set of coefficients $\{a_{jk}[\Lt_t]\}$ . 

Using equation~\eqref{eq:P_md}, we find that all relevant thermodynamic operators (the effective Hamiltonian ${K}_S$, the work and heat operators) are diagonal in the same basis of ${H}_S$. By writing each eigenvalue set as a vector, namely $\vec{k}(t) = (k_1(t), ... , k_d(t))^T$ for the effective Hamiltonian, and similarly for work and heat, we find
\begin{align}\label{eq:q-pc}
    &\vec{q}(t) = \int_0^t d\tau [\dot{F}(\tau) F^{-1}(t)]^\dag \vec{k}(\tau) \; , \\
    &\vec{w}(t) = \vec{k}(t)-\vec{q}(t) \; .
\end{align}
Since all observables commute with each other, the fluctuations of internal energy, work and heat can be simultaneously accessed when the open system undergoes phase-covariant dynamics.

\subsection{Pure decoherence} \label{sec:dephasing}
For pure decoherence dynamics, which happens whenever the interaction of the system with the environment is such that  $[{H}_S(t),{H}_I]=0$, the system can't exchange any heat with the environment, both according to the weak-coupling formulation of quantum thermodynamics and the extension through minimal dissipation -- notice, however, that different frameworks identify heat exchange with the environmental energy change, ascribing a possibly very large quantity of heat exchanged to the process \cite{Popovic2023,Picatoste2025}. 

According to the present framework, we can show that not only the average heat exchanged is zero, but also all its higher moments. This establishes, deterministically, that pure decoherence processes cannot involve any heat exchange, and that any change of internal energy of the open system can be purely assigned to work. 

To see it, notice that pure decoherence is a special case of phase covariant dynamics \cite{Chruscinski2022}. In particular, the dynamical map looks like the following
\begin{equation}
    \Phi_t [\cdot] = \sum_{j}\ket{j}\!\bra{j}\cdot \ket{j}\!\bra{j} + \sum_{j\neq k}f_{jk}(t) \ket{j}\!\bra{j}\cdot \ket{k}\!\bra{k} \; ,
\end{equation}
namely eq.~\eqref{eq:dynamical_map_pc} where $F(t)$ is the identity matrix.
Since the effective Hamiltonian is not affected by the values in $F(t)$, the Hamiltonian $K_S(t)$ for pure decoherence processes has exactly the same expression as eq.~\eqref{eq:k-pc}. 

Moreover, the fact that the heat observable is actually identically zero, that is, $O_q(t)\equiv 0$, can be seen given eq.~\eqref{eq:q-pc} for the heat observable eigenvalues, and the fact that $\dot{F}(t)\equiv 0$ for pure decoherence.
This in turn implies that the heat exchanged is identically zero deterministically (all moments are equal to zero) and that
\begin{equation}
    \braket{e^{-\beta q}} \equiv 1 \; .
\end{equation}
Moreover, it implies that the work observable can be identified with the effective Hamiltonian
\begin{equation}
    O_w(t) \equiv K_S(t) \; .
\end{equation}
Since pure decoherence maps are unital, this means that Jarzynski's equality holds at any coupling both for work and energy fluctuations:
\begin{equation}
    \braket{e^{-\beta w}} = \braket{e^{-\beta u}} = e^{-\beta \Delta \overline{F}} \; .
\end{equation}

\section{Example: qubit under phase-covariant dynamics} \label{sec:example}
In this section, we examine the behavior of the correction factor in some examples of qubit dynamics. In particular, we assume that the dynamics of the system is phase covariant, and look at both a situation compatible with the weak-coupling regime of a qubit coupled to a thermal bath (Sec.~\ref{sec:ex-wc}), and one in the strong-coupling, non-Markovian regime of a qubit coupled to a single thermal bosonic mode via a Jaynes-Cummings Hamiltonian (Sec.~\ref{sec:ex-JC}).
The details of the analytical calculations for the phase covariant maps are given in Appendix~\ref{app:phase-covariant}.

\subsection{Weak coupling}\label{sec:ex-wc}

\begin{figure}[tp!]
  \includegraphics[clip,width=\columnwidth]{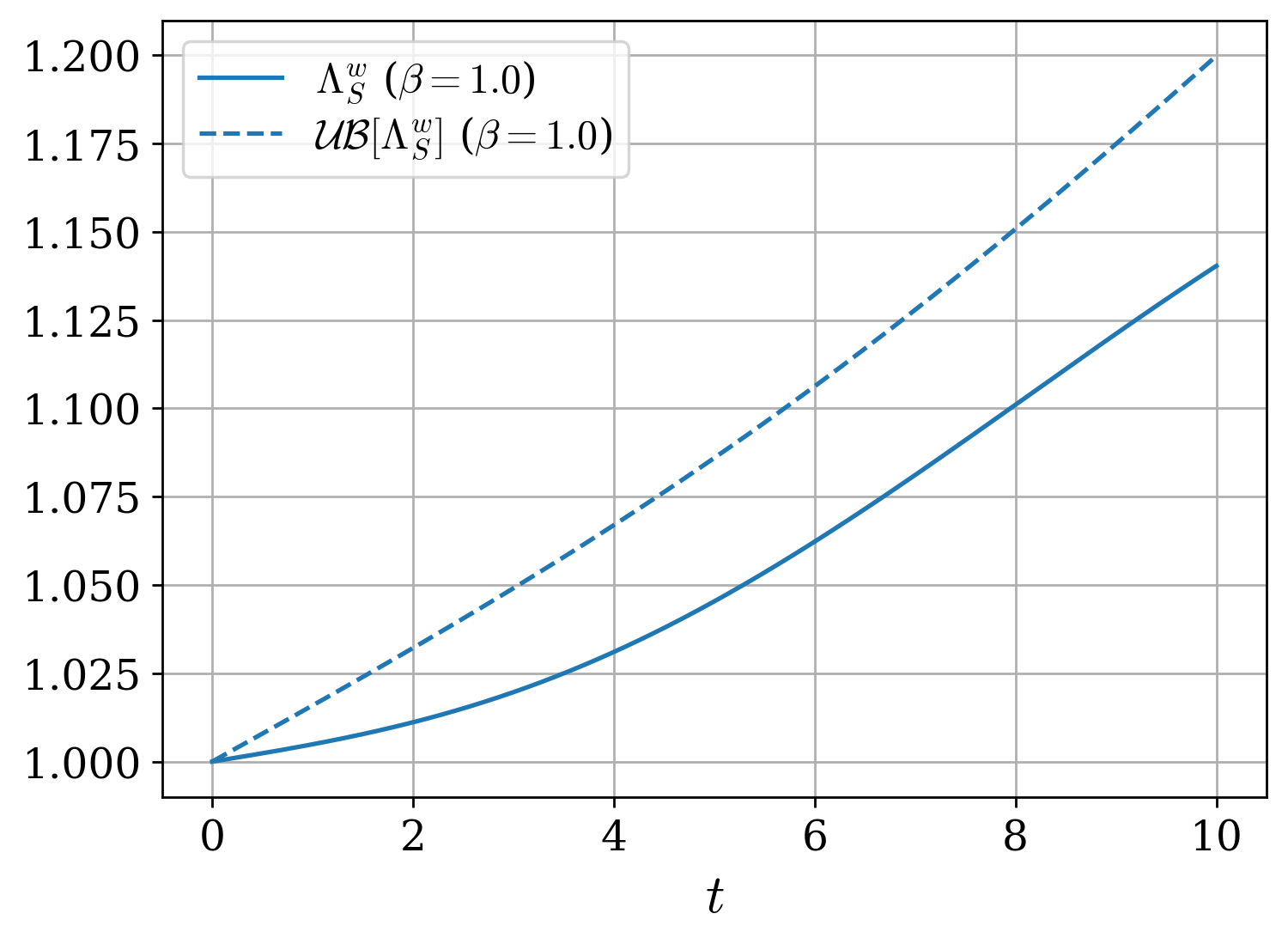}
  \includegraphics[clip,width=\columnwidth]{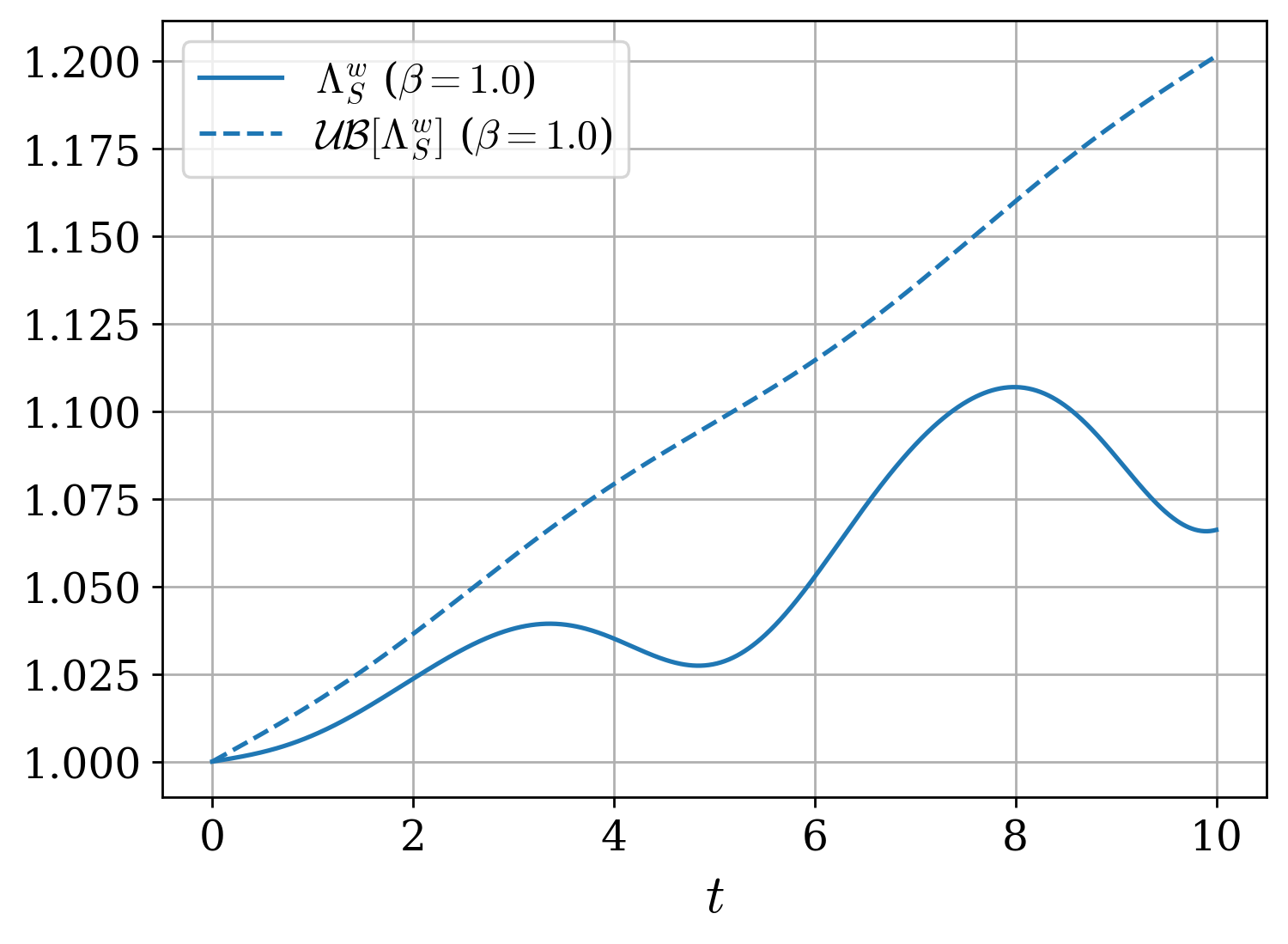}
	\caption{Temporal evolution of the correction factor $\Lambda_S^w(t)$ and its bound for (a) monotonic and (b) periodic driving protocols. Parameters are set to $\delta = 1$, $\beta = 1$, $\gamma = 0.01$, and $t_f = 10$, with all quantities expressed in units of $\omega$. For (a) we have taken $\Omega= \pi/20$, while for (b) we have taken $\Omega=\pi/5$.}
	\label{fig:wc}
\end{figure}	

\begin{figure*}[t]
  \centering
  \subfloat[]{%
    \includegraphics[width=0.45\textwidth]{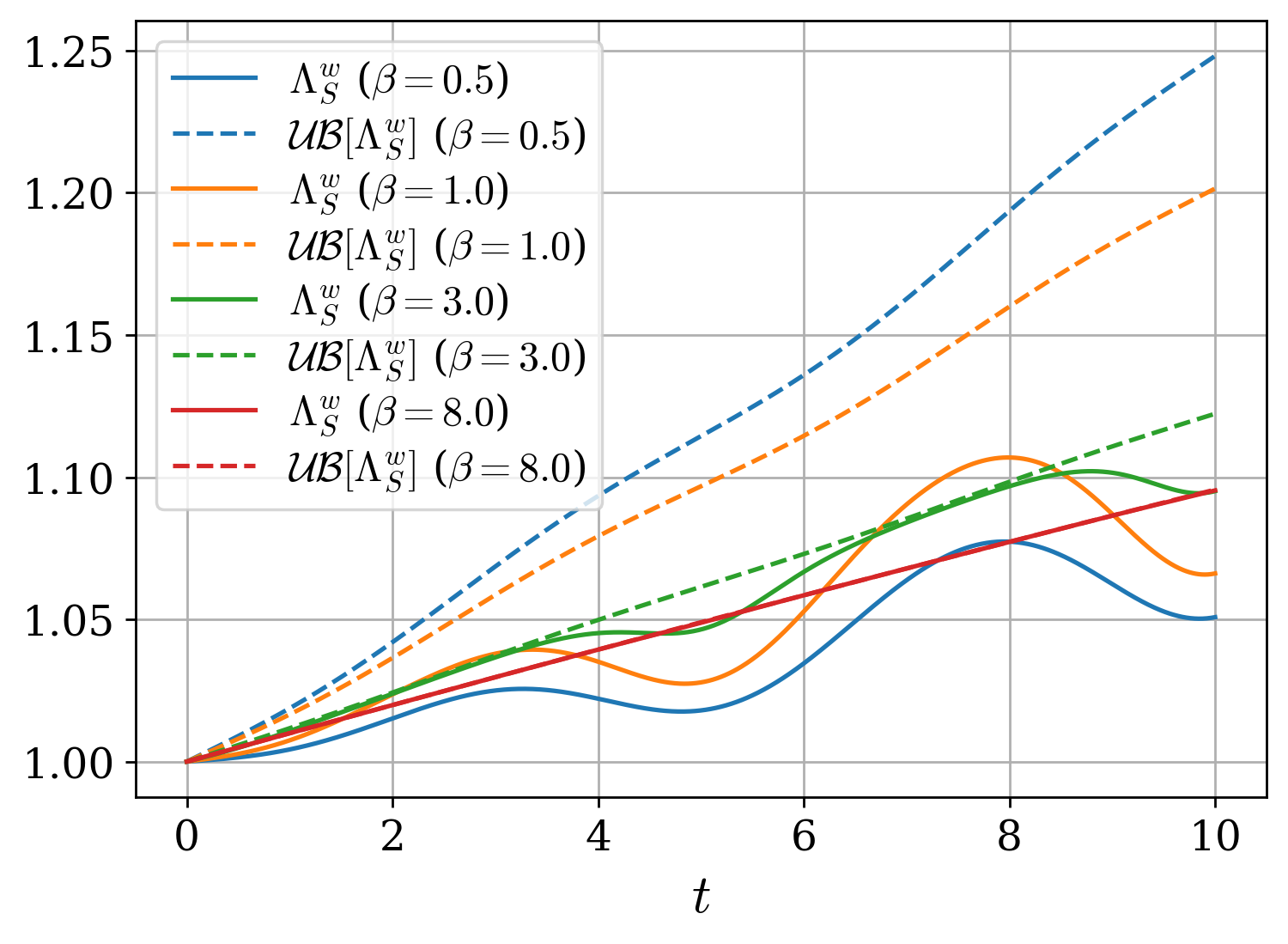}
    \label{fig:wc_temps_a}
  }
  \hfill
  \subfloat[]{%
    \includegraphics[width=0.45\textwidth]{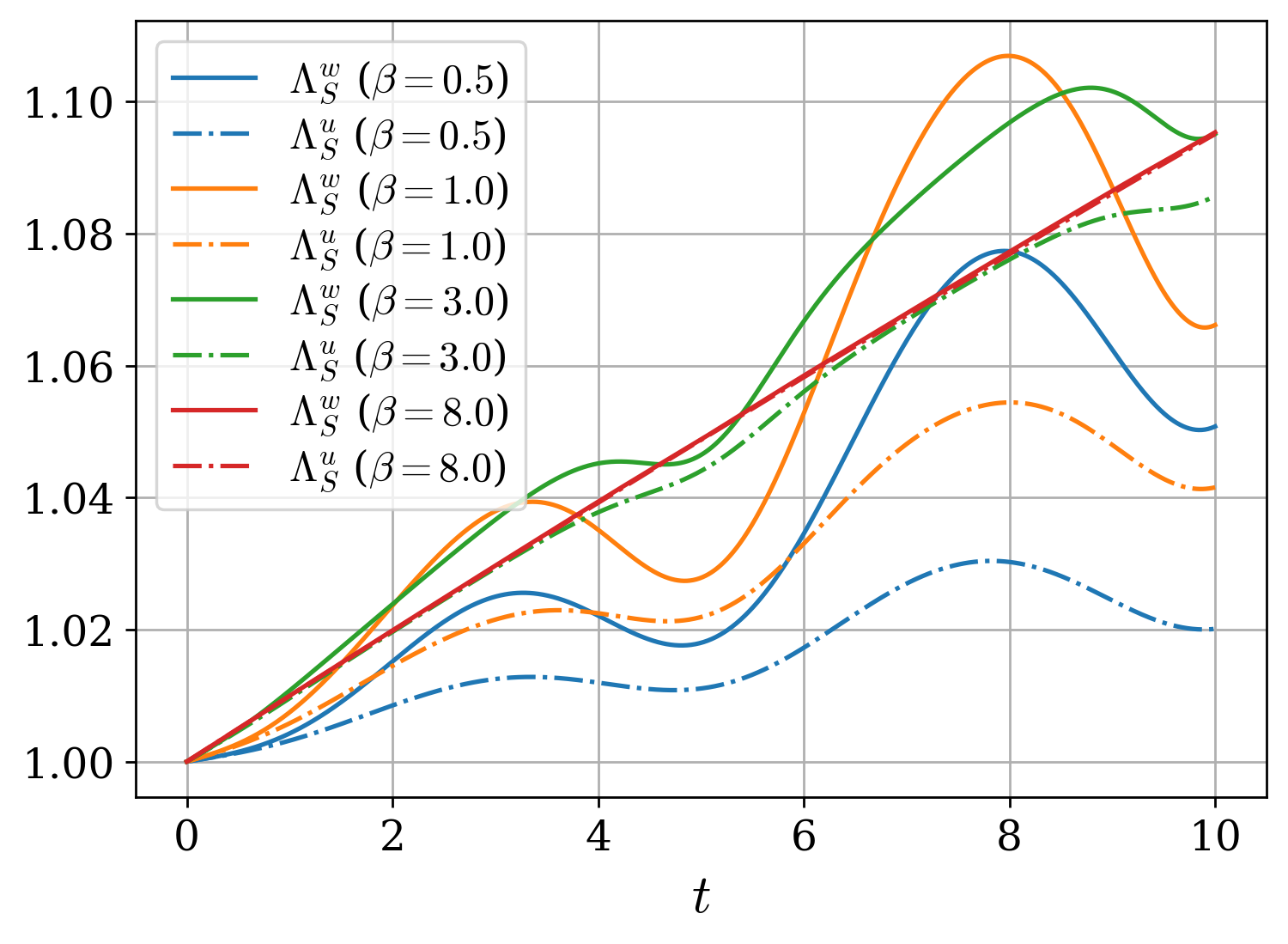}
    \label{fig:wc_temps_b}
  }
  \caption{Temperature sensitivity of the correction factors. (a) Evolution of $\Lambda_S^w(t)$ and its analytical bound for varying bath temperatures. As temperature decreases, the bound converges and the oscillations in the exact factor flatten. (b) Comparison of $\Lambda_S^w(t)$ with the internal energy correction factor $\Lambda_S^u(t)$ for the same temperature range. Parameters in both figures are: $\delta = 1$, $\gamma = 0.01$, and $t_f = 10$ in units of $\omega$.}
  \label{fig:wc_lambda_temps}
\end{figure*}

Let us assume that our system of interest is a qubit that is weakly coupled to a large white noise thermal bath at inverse temperature $\beta$. We assume that the driving protocol performed on the system is described by a Hamiltonian $H_S(t)= \omega(t)\sigma_z/2$, where $\omega(t)$ denotes the splitting of the driven qubit. Then, we assume that the dynamics of the system is well approximated by the following master equation:
\begin{align}\nonumber
\dot{\rho}(t) =& -i [H_S(t), \rho(t)] + \gamma_+ \mathcal{D}_{\sigma_+}[\rho(t)] + \gamma_- \mathcal{D}_{\sigma_-}[\rho(t)] \\&+ \gamma_z \mathcal{D}_{\sigma_z}[\rho(t)] \; ,
\end{align}
where the following shorthand notation is used to indicate dissipator contributions
\begin{align}
\mathcal{D}_{L}[\rho(t)]= L\rho(t)L^\dag -\frac{1}{2}\{L^\dag L, \rho(t)\} \; ;
\end{align}
further notice that any Lamb shift contribution to the Hamiltonian has been neglected \cite{gardiner2000}. The rates are given by $\gamma_-= \gamma (N+1)$ and $\gamma_+= \gamma N$, where we take $N=(e^{\beta\omega(0)}-1)^{-1}$ and $\gamma$ is a small coupling parameter. The value of $\gamma_z$ does not influence our results.

We look at two distinct cases of frequency driving. The first is given by 
\begin{equation}\label{eq:driving_shape}
\omega(t)= \omega +\delta \sin^2(\Omega t) \; ,
\end{equation}
where the driving protocol is assumed to last until a time $t_f= \pi / 2\Omega$, corresponding to the first peak of the driving, and thus describing a monotonic ramping of the qubit frequency gap from $\omega$ to $\omega +\delta$. Other types of ramps do not influence significantly the qualitative behavior of the correction factor $\Lambda_S^w(t)$. The second case we consider, therefore, is non-monotonic driving, where the energy splitting change eq.~\eqref{eq:driving_shape} is implemented with a higher frequency such that $t_f= 2 \pi / \Omega$.

For both situations, it is important to remark that it only makes sense to consider short driving times $\gamma t_f \ll 1$. After that, the system begins to equilibrate and the magnitude of the inverse dynamical map becomes large. This, as explained in Sec.~\ref{sec:work-heat-wc}, leads to fast divergence of the correction factor as well as to a near-equilibrium state at $t_f$, which is outside of typical fluctuation relations scenarios.

We can see the behavior of the correction factor $\Lambda_S^w(t)$, from eq.~\eqref{eq:lambda_w}, in time in Fig.~\ref{fig:wc}, for both shapes of driving. For the short times considered, the factor remains quite close to the closed-system value of one, however increasingly parting from it with time. While the monotonic driving leads to a monotonically increasing factor, the periodic driving leads to oscillatory behavior in $\Lambda_S^w(t)$. In both cases, the upper bound eq.~\eqref{eq:lambdaw_wc_bound} on the correction factor follows the trend of the exact quantity, but shows a monotonic behavior also for the case of periodic driving.

The factors $\Lambda_S^w(t)$ and their bounds are sensitive to the bath temperature. The bound for $\Lambda_S^w(t)$ decreases for lower temperatures and converges from above to a final curve. Moreover, for low temperatures the oscillations in the factor $\Lambda_S^w(t)$ flatten out towards their bounds. This can be seen from Fig.~\ref{fig:wc_temps_a}, as well as understood analytically, see Appendix~\ref{app:phase-covariant}. In Fig.~\ref{fig:wc_temps_b} we also show, for the same temperatures, the behavior of the factor $\Lambda_S^u(t)$  (eq.~\eqref{eq:lambda_u}) for internal energy fluctuations. It shows similar but slightly less pronounced features with respect to $\Lambda_S^w(t)$. In particular, it still increases with time for the considered window. Differently from $\Lambda_S^w(t)$, however, it will not increase exponentially when closer to equilibration timescales; this is due to the fact that it does not contain the inverse map.

\subsection{Autonomous non-Markovian system: the Jaynes-Cummings model}\label{sec:ex-JC}
\begin{figure*}[t]
  \centering
  \subfloat[]{%
    \includegraphics[width=0.45\textwidth]{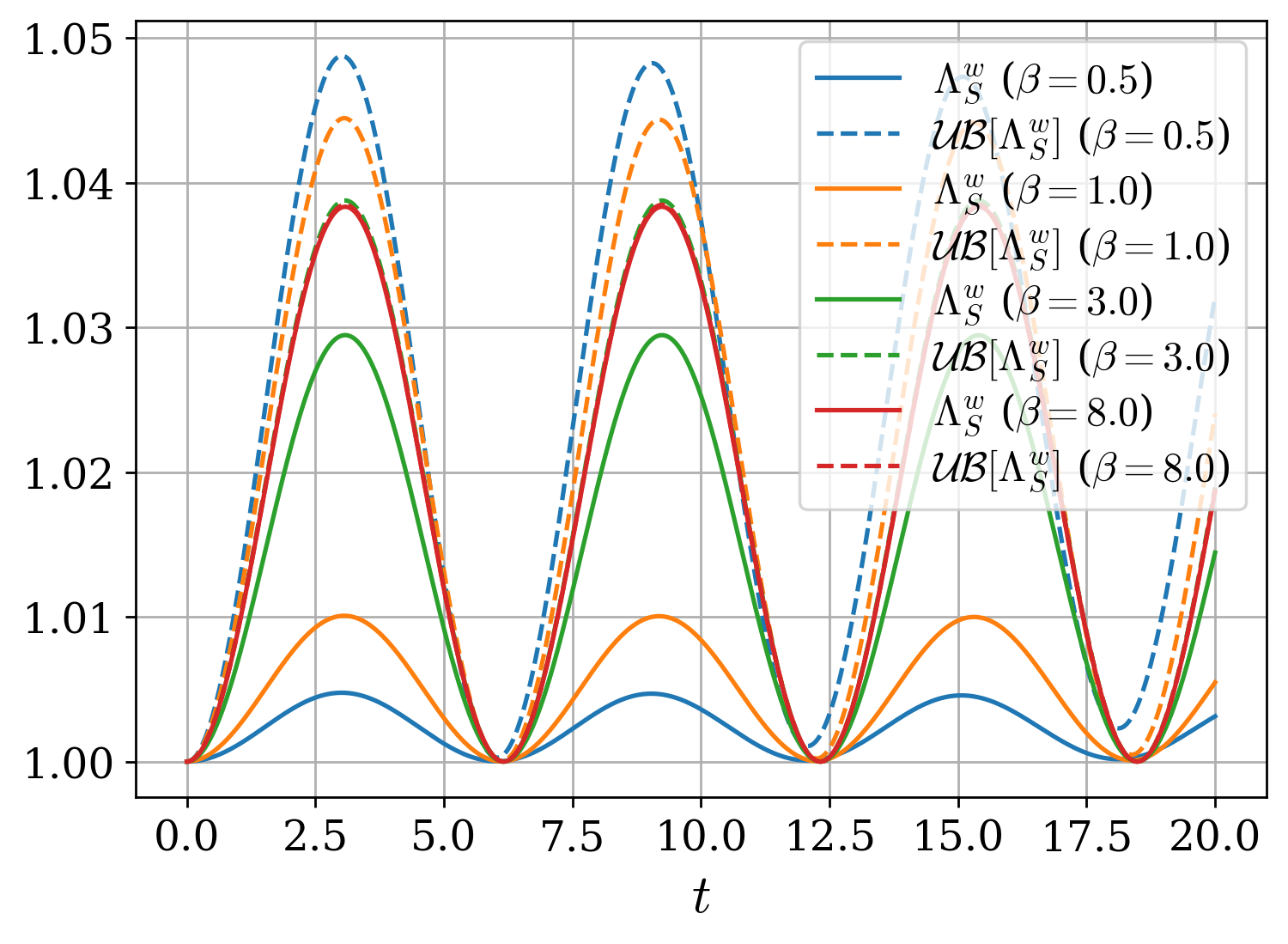}
    \label{fig:JC_lowT_a}
  }
  \hfill
  \subfloat[]{%
    \includegraphics[width=0.45\textwidth]{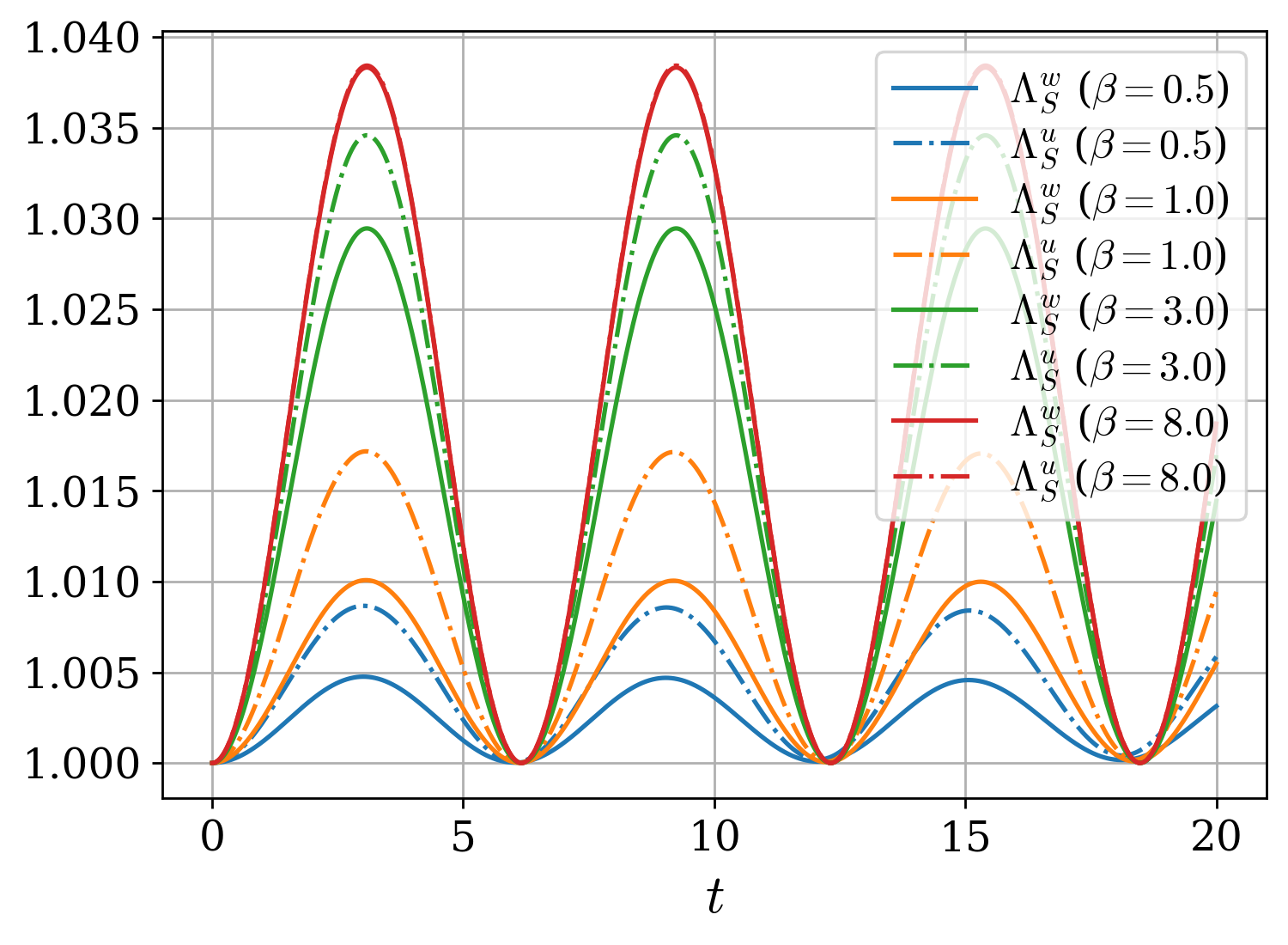}
    \label{fig:JC_lowT_b}
  }
  \caption{Correction factors at different temperatures for the Jaynes-Cummings model. (a) Work correction factor $\Lambda_S^w(t)$ for varying temperatures; as temperature decreases, the factor rises to saturate the analytical bound. (b) Comparison with the internal energy factor $\Lambda_S^u(t)$, which remains consistently higher for this parameter set. The finite environment induces distinct, clean oscillations and recurrences. Parameters are in units of $\omega$, and are given by $\omega_m = 2$, and $g = 0.01$.
  }
	\label{fig:JC_lowtT}
\end{figure*}

Let us now assume that the total system is actually isolated, i.e., that no external driving is performed on the qubit. We instead assume that the latter is coupled to a single bosonic mode via linear, Jaynes-Cummings interaction:
\begin{align}H_{SE} = \frac{\omega}{2}\sigma_z + \omega_m a^\dag a + g (\sigma_+ a + \sigma_- a^\dag) \; .
\end{align}
When the mode is initialized in a thermal state, the reduced, time-local master equation for the qubit is phase-covariant and reads \cite{Smirne2010}:
\begin{align}\nonumber
\dot{\rho}(t) =& -i \left[\frac{\omega(t)}{2}\sigma_z, \rho(t)\right] + \gamma_+(t) \mathcal{D}_{\sigma_+}[\rho(t)] \\&+ \gamma_-(t) \mathcal{D}_{\sigma_-}[\rho(t)] + \gamma_z(t) \mathcal{D}_{\sigma_z}[\rho(t)] \; . \label{eq:JC_me}
\end{align}
Namely, tracing out the bosonic degree of freedom leads to an emergent driving of the system frequency in a completely autonomous way (without external driving). This effect has been measured experimentally in the same model \cite{Colla2024exp}. 

Because the environment is not a continuum of bosonic modes and in general not weakly coupled to the qubit, this model represents a paradigmatic case of a non-Markovian, strongly coupled dynamics. Indeed, the exact dynamics has been shown to break P-divisibility, which implies the presence of information backflow from the mode to the system \cite{Wissmann2015,Teittinen2018,Theret2025}. In this case, the minimal dissipation framework is required to describe the thermodynamic properties of the qubit without requiring access to the environmental mode. 

Another crucial difference from the example considered in Sec.~\ref{sec:ex-wc}, which is again due to the finiteness of the single bosonic mode as an environment, is that the system does not evolve towards equilibrium. For this reason, the inverse of the dynamical map is typically finite also at longer timescales. However, the master equation \eqref{eq:JC_me} can fail to exist at singular points in time precisely due to instantaneous non-invertibility of the map \cite{Andersson2007,Chruscinski2018}, an instance that typically only happens asymptotically for dynamics such as the one described in Sec.~\ref{sec:ex-wc}. At these discrete points in time, the inverse map does not exist, which implies that our defined operator $\mathfrak{P}(t)$ (eq.~\eqref{eq:P_md}) also no longer exists at those times. There may also be instances where the inverse map is still well defined but very large in magnitude around discrete times: these cases will give rise to isolated ``spikes'' in the behavior of the factor $\Lambda_S^w(t)$, which would not appear, for example, in the factor $\Lambda_S^u(t)$. Because such times appear mostly close to the resonant case $\omega \approx \omega_m$, we will consider here only non-resonant scenarios.

\begin{figure}[tp!]
  \includegraphics[clip,width=\columnwidth]{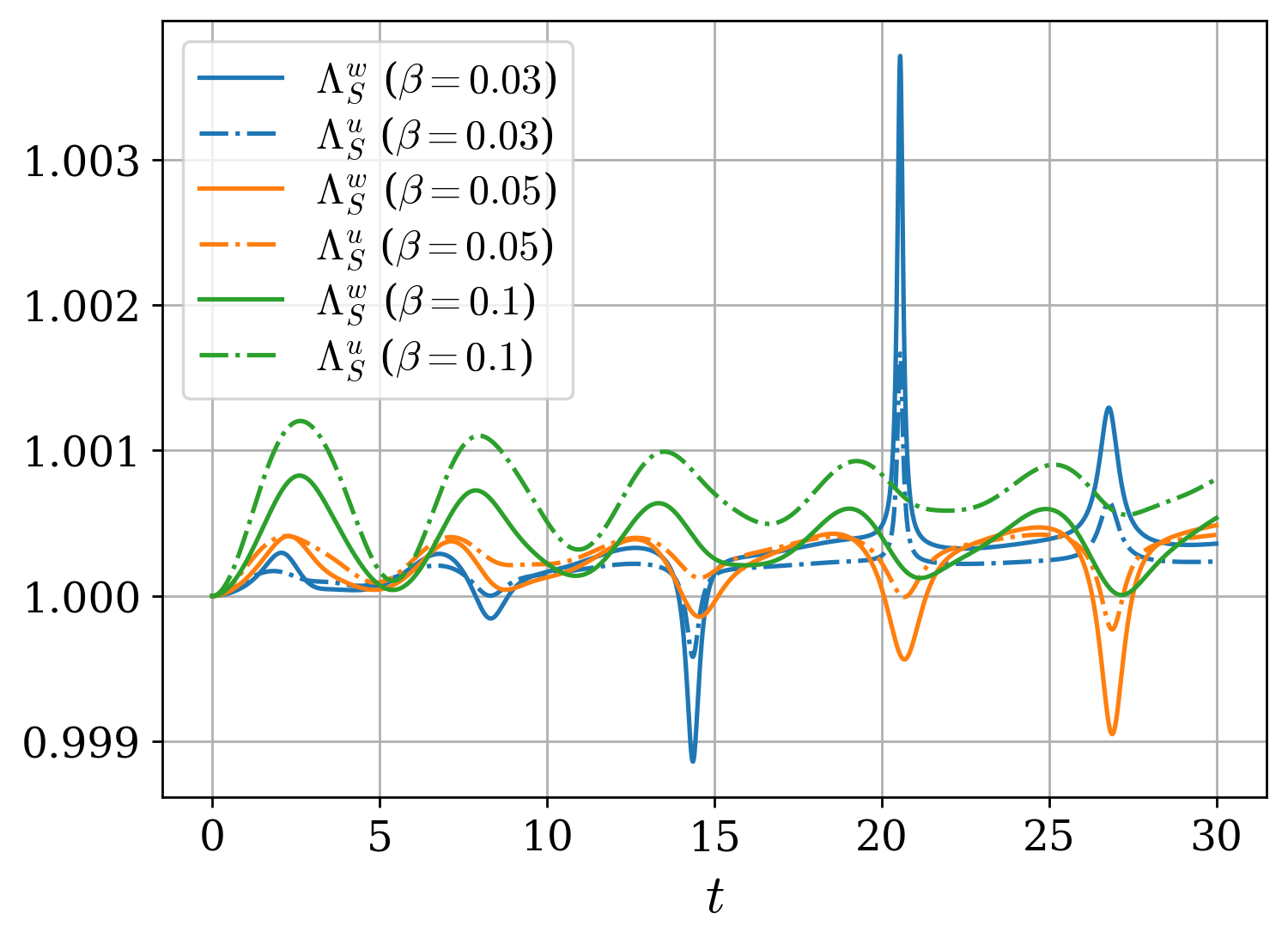}
	\caption{High-temperature dynamics of the correction factors $\Lambda_S^w(t)$ and $\Lambda_S^u(t)$ for the Jaynes-Cummings model. For both factors, the regular oscillations observed at low temperature are replaced by complex, multi-frequency behavior and sharp peaks. The factor can temporarily drop below 1. Parameters are in units of $\omega$ and given by $\omega_m = 2$, and $g = 0.01$.}
	\label{fig:JC_highT}
\end{figure}	

\begin{figure*}[t]
  \centering
  \subfloat[]{%
    \includegraphics[width=0.45\textwidth]{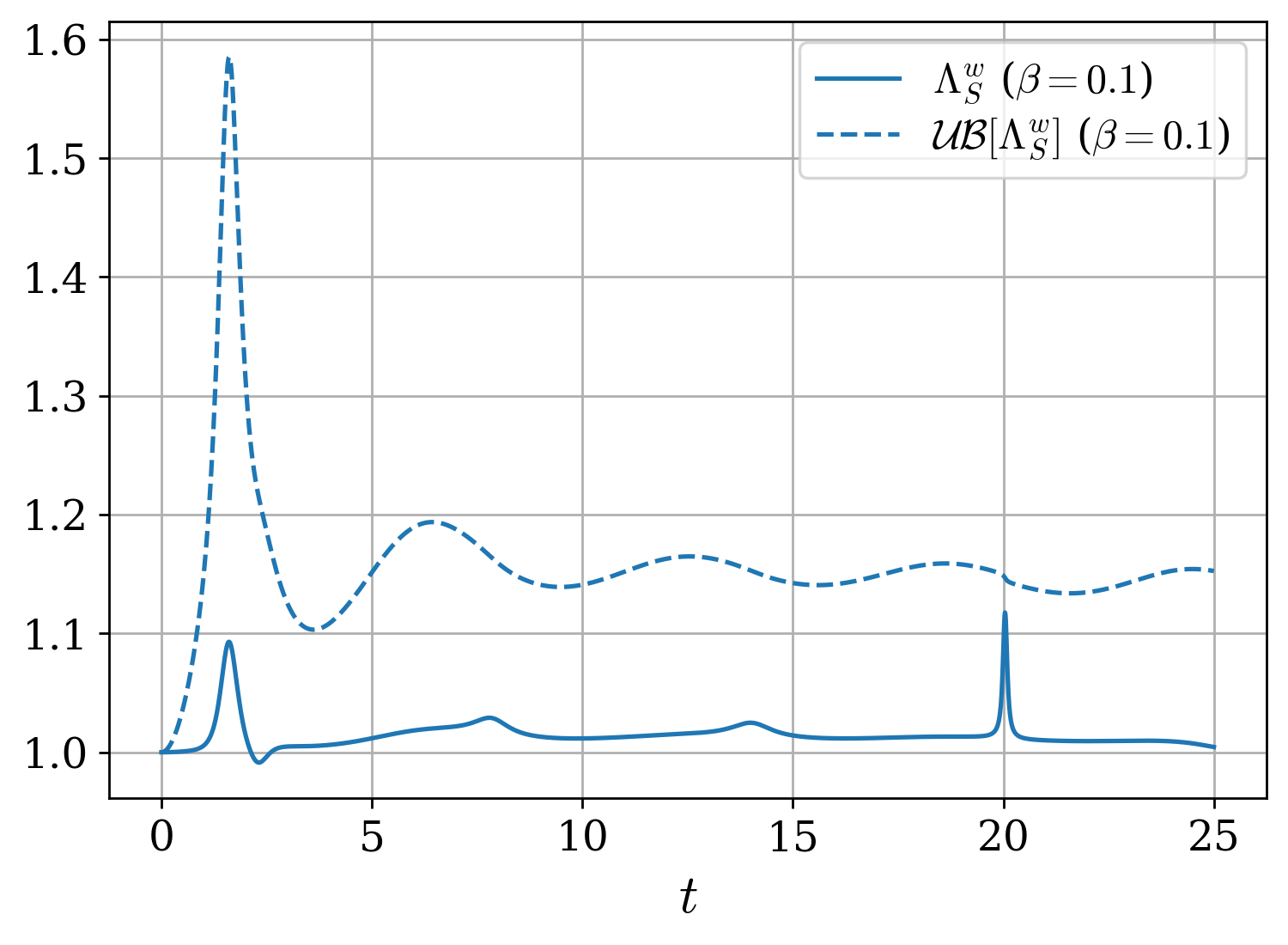}
    \label{fig:JC_ECG_a}
  }
  \hfill
  \subfloat[]{%
    \includegraphics[width=0.45\textwidth]{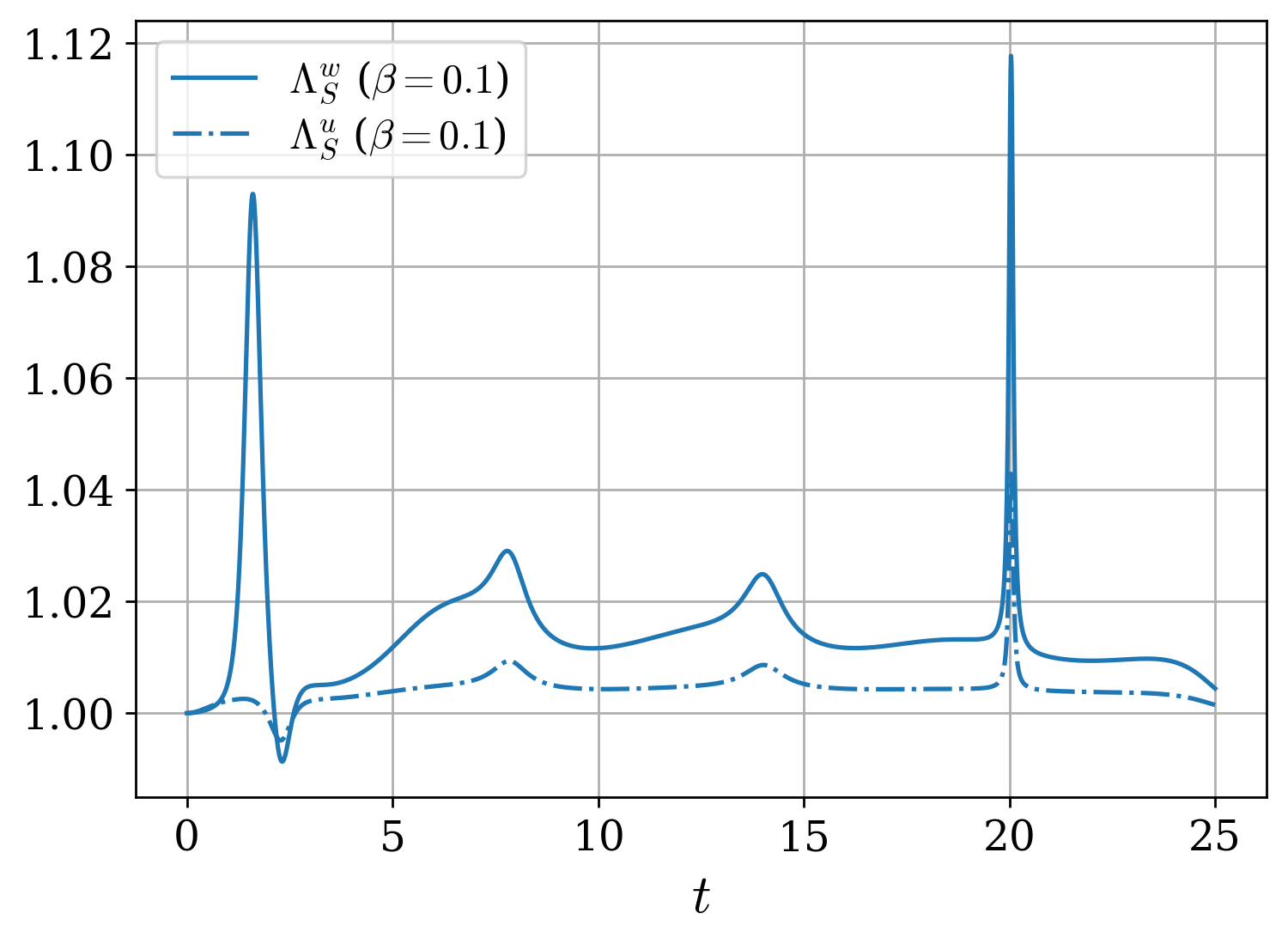}
    \label{fig:JC_ECG_b}
  }
  \caption{Correction factors at stronger couplings. (a) Evolution of $\Lambda_S^w(t)$ and its bound showing a highly structured profile with higher deviations from unity. (b) Comparison between $\Lambda_S^w(t)$ and $\Lambda_S^u(t)$. While both exhibit sharp, isolated peaks and temporary values below one, the work correction factor $\Lambda_S^w(t)$ displays a characteristic initial peak absent in the internal energy factor $\Lambda_S^u(t)$. Parameters are in units of $\omega$ and given by $\omega_m = 2$, and $g=0.1$.
  }
	\label{fig:JC_ECG}
\end{figure*}

In Fig.~\ref{fig:JC_lowtT}, we show the behavior of the factor $\Lambda^w_S(t)$ for relatively low coupling strength, at different temperatures. Similarly to the weak-coupling case of the previous section (Fig.~\ref{fig:wc_lambda_temps}), at lower temperatures the bounds converge from above to a single line, while the factor $\Lambda^w_S(t)$ rises closer to saturate the bound, Fig.~\ref{fig:JC_lowT_a}. The behavior of the internal energy fluctuation parameter $\Lambda^u_S(t)$ is also very similar, see Fig.~\ref{fig:JC_lowT_b}. In this particular set of parameters, it is consistently higher than $\Lambda^w_S(t)$, but we note that this is not a general feature and depends highly on the parameter combinations, including temperature, and the detuning between the system and the mode. What changes drastically from Fig.~\ref{fig:wc_lambda_temps}, instead, is the much cleaner oscillatory behavior; while for longer times the minima may no longer reach the value one, the maximum value of the factors is given by the peaks visible already at short timescales in Fig.~\ref{fig:JC_lowtT}. This, as mentioned before, is due to the finite size of the environment and its associated recurrences.

At higher temperatures, the qualitative behavior of $\Lambda^w_S(t)$ changes, with the oscillations disappearing progressively with rising temperature, see Fig.~\ref{fig:JC_highT}. The behavior is however less regular: more frequencies are involved and there is the presence of sharp peaks at distinct points in time. This feature, and the behavior in time in general, is shared also by the internal energy fluctuations factor $\Lambda^u_S(t)$. Notably, in these parameter regimes both factors can become temporarily lower than one. The bounds for $\Lambda^w_S(t)$ are however not tight, with deviations from one that are about one order of magnitude larger than the deviations of $\Lambda^w_S(t)$.

Lastly, we show the effects of both a stronger coupling parameter $g$, and of a smaller detuning $\Delta$ between system and mode, see Fig.~\ref{fig:JC_ECG}. For the Jaynes-Cummings model, this regime corresponds to ultrastrong coupling \cite{Forn-Diaz2019}. The deviation of the factor $\Lambda^w_S(t)$, as well as its bound, from the value of one is now higher -- about one to two orders of magnitude greater than what observed in the parameter regime of Fig.~\ref{fig:JC_highT}. The shape of the factor, as well as of $\Lambda^u_S(t)$ (see their comparison in Fig.~\ref{fig:JC_ECG_b}), is much more structured than in previous examples. They both temporarily can become lower than one, but also feature very sharp isolated peaks. Notice, however, the appearance of an initial peak in $\Lambda^w_S(t)$ which is instead absent in $\Lambda^u_S(t)$.

\section{Conclusions}

In this work, we have proposed a method to evaluate general thermodynamic fluctuations in open quantum systems, based on performing a two-point measurement scheme on the system using suitable thermodynamic observables. Crucially, this method allows one to access fluctuations of quantities that depend also on the intermediate non-equilibrium evolution of the system (such as work and heat in a general open quantum system dynamics) and can be carried out accessing only the reduced system degrees of freedom. 

The ``heavy-lifting'' of this method is done by using the explicit dynamical map for the system in the definition of the thermodynamic two-point observables. While this is in general a complicated object, it can be accessed with adequate control of the system degrees of freedom via quantum process tomography. This allows us to compute fluctuations for virtually any open quantum system quantity, with a few caveats:
\begin{itemize}
    \item because we employ the inverse of the dynamical map, the fluctuations can become very large when the inverse becomes very large (close to non-invertibility times) and may even fail to exist at a time where the dynamical map is not invertible;
    \item fluctuations for the same average quantity may yield different outcomes depending on the exact definition of the observable used in the TPMS. This implies that the fluctuations are measurement-protocol dependent. 
\end{itemize}
On the other hand, the freedom of choice in the measurement protocol also allows one to extend the framework to non-diagonal initial states (as long as the system eigenbasis is known), avoiding a common limitation of the standard two-point measurement scheme.
Moreover, the flexibility of this method allowed us to recover both the standard TPMS scheme for work in closed systems \cite{Talkner2007a}, and the work operator approach proposed in \cite{Allahverdyan2005} in appropriate limiting cases. This feature highlights the unifying spirit of our proposal.

Applied in the weak-coupling regime to internal energy and work operators, our framework includes and  generalizes what is proposed in \cite{Goold2021} to obtain work fluctuations (rather than just internal energy). For both thermodynamic quantities, it provides an equality that reverts to Jarzynski's in the limit of closed systems, but with a correction factor that takes into account the non-unitality of the dynamical map as well as the effect of dissipating some of the energy variation into heat.

The method is also straightforwardly extended to include the study of arbitrary, non-Markovian, and strongly coupled open quantum systems. This can be done by adopting the framework of minimal dissipation for the definition of thermodynamic quantities in these regimes, which then include renormalization effects due to the structured interaction with the environment. 

We have investigated the effect of specific dynamical classes and symmetries for the open system on the fluctuation relations. In particular, we have shown that the pure decoherence case is particularly special as it deterministically does not contain any heat contribution; as a consequence of this and the inherent unitality of pure decoherence maps, it constitutes a class of open system dynamics for which the Jarzynski equality for work fluctuations is identically true at any coupling strength. We have also investigated phase-covariant dynamics, for which all two-point fluctuations of thermodynamic quantities of interest (internal energy, work and heat) can be simultaneously accessed.

As a particular case, we have looked explicitly at the shape and size of the correction factors to Jarzynski's equality for a qubit undergoing phase-covariant dynamics, for which we have derived analytical expressions. Moreover, this kind of dynamics can both describe an externally driven qubit weakly coupled to a Markovian bath, as well as a qubit coupled to a non-Markovian, finite bath which is responsible for emergent driving and work. In both situations, the correction factor depends on time and only slightly deviates from one (unless the dynamical map is close to be non-invertible).

Although we presented purely analytical results, the framework can be implemented numerically with the help of simulation techniques for open quantum system dynamics -- such as HEOM, tensor network methods, or pseudomodes \cite{Xu2026}-- in order to study thermodynamic fluctuations for more complex systems and construct the observables for relevant path-dependent quantities. 

Lastly, while this work has focused mainly on Jarzynski-like types of fluctuation relations, the approach we introduced here can be used to look at all moments of heat and work, with implications, e.g. on thermodynamic uncertainty relations.

\acknowledgments
The authors would like to thank Giacomo Guarnieri for the careful reading of the manuscript.
This work has been supported from Ministero dell’Università e della Ricerca (MUR) and Next Generation EU via the NQSTI-Spoke1-BaC project QuSynKrono (contract n. PE00000023-QuSynKrono) and via the PRIN 2022 project Quantum Reservoir Computing (QuReCo) (contract n. 2022FEXLYB). A.C. acknowledges support by the DFG funded Research Training Group ``Dynamics of Controlled Atomic and Molecular Systems'' (RTG 2717).

\bibliography{biblio_fluctuations}

\appendix
\section{Phase-covariant dynamics}\label{app:phase-covariant}

\subsection{Explicit dynamics in terms of the generator parameters}\label{app:dynamics}
Let us assume a phase-covariant generator for the qubit:
\begin{align}\nonumber
    \Lt_t[\cdot] = -i \left[\frac{\omega(t)}{2} \sigma_z, \cdot \right] + & \gamma_+(t)\mathcal{D}_{\sigma_+}[\cdot] + \gamma_-(t)\mathcal{D}_{\sigma_-}[\cdot] \\ & +  \gamma_z(t)\mathcal{D}_{\sigma_z}[\cdot] \; ,
\end{align}
where we therefore assume that the Hamiltonian associated to the system is given by $H_S(t)=\frac{\omega(t)}{2} \sigma_z$.
From the coefficients $\omega$,$\gamma_{+,-,z}$ one can write the exact dynamical map acting on the basis $\{\mathbb{I}, \sigma_x, \sigma_y, \sigma_z\}$ as the following matrix \cite{Smirne2016}
\begin{equation}
    \Phi_t = \begin{bmatrix}
    1&0&0&0 \\
    0& a(t) &-b(t) & 0 \\
    0& b(t) &a(t) & 0 \\
    c(t) & 0 & 0 & d_{\parallel}(t)
    \end{bmatrix} \; .
\end{equation}
In the above, we have defined $\kappa(t) = \gamma_+(t) + \gamma_-(t)$ and $\xi(t) = \gamma_+(t) - \gamma_-(t)$ and the integrals
\begin{align}
    I(t) =& \int_0^t ds \kappa(s) \\
    J(t) =& \int_0^t ds\xi(t) e^{I(s)} \\
    d_{\perp}(t) =& e^{-I(t)/2}e^{-2 \int_0^t ds \gamma_z(s)} \; ,
\end{align}
so that the coefficients of the dynamical map read
\begin{align}
    a(t) =& d_{\perp}(t)\cos\left(\int_0^t ds \omega(s)\right) \\
    b(t) =& d_{\perp}(t)\sin\left(\int_0^t ds \omega(s)\right) \\
    c(t) =&  J(t) e^{-I(t)}\\
    d_{\parallel}(t) =& e^{-I(t)}\; .
\end{align}
We can then find its derivative and inverse:
\begin{equation}
    \dot{\Phi}_t = \begin{bmatrix}
    0&0&0&0 \\
    0& \dot{a}(t) &-\dot{b}(t) & 0 \\
    0& \dot{b}(t) & \dot{a}(t) & 0 \\
    \dot{c}(t) & 0 & 0 & \dot{d}_{\parallel}(t) 
    \end{bmatrix}\; ,
\end{equation}
\begin{equation}
    \Phi^{-1}_t = \begin{bmatrix}
    1&0&0&0 \\
    0& a(t)/d_{\perp}^2(t) &b(t)/d_{\perp}^2(t) & 0 \\
    0& -b(t)/d_{\perp}^2(t) &a(t)/d_{\perp}^2(t) & 0 \\
    -c(t)/d_{\parallel}(t) & 0 & 0 & 1/d_{\parallel}(t)
    \end{bmatrix}\; .
\end{equation}
Because of this structure, and recalling that 
\begin{equation}
   \mathfrak{P}(t) =  \int_0^t d\tau \left(\Phi_t^{-1}\right)^\dag\left(\dot{\Phi}_\tau\right)^\dag [H_S(\tau)] \; ,
\end{equation}
we find
\begin{equation}
   \mathfrak{P}(t) =  \mathfrak{P}_0(t)\mathbb{I} +\mathfrak{P}_3(t)\sigma_z \; ,
\end{equation}
where 
\begin{align}
    \mathfrak{P}_0(t) &= \frac{1}{2} \int_0^t d\tau \omega(\tau) \left( \dot{c}(\tau) - c(t) \frac{\dot{d}_{||}(\tau)}{d_{||}(t)}\right)  \; ,\\
    \mathfrak{P}_3(t) &= \frac{1}{2} \int_0^t d\tau \omega(\tau)  \frac{\dot{d}_{||}(\tau)}{d_{||}(t)} \; .
\end{align}
By defining the following integrals:
\begin{align}
    T_A(t) &= \int_0^t d\tau \omega(\tau) \kappa(\tau)J(\tau)e^{-I(\tau)}  \; ,\\
    T_B(t) &= \int_0^t d\tau \omega(\tau) \kappa(\tau)e^{-I(\tau)} \; , \\
    T_C(t) &= \int_0^t d\tau \omega(\tau) \xi(\tau)\; ,
\end{align}
we find explicitly
\begin{align}
    \mathfrak{P}_0(t) &= \frac{1}{2} \left[ T_C(t) - T_A(t) +J(t)T_B(t)\right]  \; ,\\
    \mathfrak{P}_3(t) &= -\frac{1}{2} T_B(t)e^{I(t)} \; .
\end{align}
From the above we can write the work operator
\begin{equation}
    O_w(t) = W_0(t) \mathbb{I} + W_3(t) \sigma_z
\end{equation}
with
\begin{align}
    W_0(t) =& -\mathfrak{P}_0(t) \; ,\\
    W_3(t) =& \frac{\omega(t)}{2}- \mathfrak{P}_3(t)\; .
\end{align}
Then, using $\Phi_t[\mathbb{I}]=\mathbb{I} +c(t)\sigma_z$, the coefficient for the work fluctuations reads
\begin{align}\nonumber
    \Lambda_S^w(t) = & \frac{e^{-\beta W_0(t)}}{\cosh\left(\beta \omega(t)/2\right)} \times\\ & \quad\left[ \cosh(\beta W_3(t)) - c(t)\sinh(\beta W_3(t))\right] \; . \label{eq:lambdaw_pc}
\end{align}

\subsection{Bounds on correction factors}\label{app:bounds}
 
 Again using $\Phi_t[\mathbb{I}]=\mathbb{I} +c(t)\sigma_z$, we obtain an expression for the eigenvalue $\lambda_{\mathrm{max}}\{\Phi_t[\mathbb{I}]\}$ appearing in the bounds on work and internal energy fluctuations due to the non-unitality of the map:
\begin{equation}
    \lambda_{\mathrm{max}}\{\Phi_t[\mathbb{I}]\} = 1+ |J(t)|e^{-I(t)} \; . 
\end{equation}
Similarly, the eigenvalue appearing in the bound on work fluctuations due to heat exchange reads
\begin{equation}
    \lambda_{\mathrm{max}}\{\mathfrak{P}(t)\} = \mathfrak{P}_0(t)+ |\mathfrak{P}_3(t)| \; ,
\end{equation}
so that the bound on $\Lambda_S^w(t) $ reads
\begin{equation}
    \Lambda_S^w(t)  \leqslant \left[ 1+ |J(t)|e^{-I(t)}\right] e^{\beta (\mathfrak{P}_0(t) + |\mathfrak{P}_3(t)|)} \; .\label{eq:lambdaw_pc_bound}
\end{equation}

Similarly, we write the lower bound on dissipated work
\begin{equation}
    \braket{w}-\Delta \overline{F} \geqslant - \mathfrak{P}_0(t) - |\mathfrak{P}_3(t)| -\frac{1}{\beta}\ln\left(1+ |J(t)|e^{-I(t)}\right) \; .
\end{equation}
Notice that for unital maps $J(t)=\mathfrak{P}_0(t)=0$, so the bounds read
\begin{align}
& \Lambda_S^w(t)  \leqslant e^{\beta |\mathfrak{P}_3(t)|} \;, \\
&\braket{w}-\Delta \overline{F} \geqslant  - |\mathfrak{P}_3(t)|  \; .
\end{align}

The exact dissipated work can be calculated knowing that the initial system state is $\rho_S(0) = [\mathbb{I} + v_z\sigma_z]/2$ with $v_z=\tanh(-\beta \omega(0)/{2})$, which gives
\begin{equation}
\braket{w}  = \left[v_z+J(t)\right] e^{-I(t)}W_3(t) + W_0(t) - \frac{\omega(0)}{2}v_z 
\end{equation}
and knowing
\begin{equation}
\Delta \overline{F}  = -\frac{1}{\beta}\ln \left[\cosh\left(\frac{\beta \omega(t)}{2}\right)\right] + \frac{1}{\beta}\ln \left[\cosh\left(\frac{\beta \omega(0)}{2}\right)\right].
\end{equation}

Using the above expressions one can also show how the coefficient $ \Lambda_S^w(t)$ approaches its bound in the weak-coupling case of Sec.~\ref{sec:ex-wc} for small temperatures. Indeed, in this case both rates $\gamma_{\pm}$ are always positive and such that $\xi(t) < 0$. This implies that $c(t)<0$, that $\mathfrak{P}_3(t)<0$ and in turn that $W_3(t)= \omega(t)/2+|\mathfrak{P}_3(t)|>0$. For large $\beta$, we can approximate $\sinh(\beta W_3(t))\approx \cosh(\beta W_3(t)) \approx e^{\beta W_3(t)}/2$ in eq.~\eqref{eq:lambdaw_pc}, with which we see that the bound \eqref{eq:lambdaw_pc_bound} is saturated.

\end{document}